\documentclass[]{{IEEEtran}}
\IEEEoverridecommandlockouts
\usepackage{amsmath,amsfonts}
\usepackage[ruled,vlined]{algorithm2e}
\usepackage{array}
\usepackage[caption=false,font=normalsize,labelfont=sf,textfont=sf]{subfig}
\usepackage{textcomp}
\usepackage{stfloats}
\usepackage{url}
\usepackage{verbatim}
\usepackage{graphicx}
\usepackage{cite}
\usepackage{xcolor}
\usepackage[nolist,nohyperlinks]{acronym} 
\usepackage{amsmath,amssymb,amsfonts}
\usepackage{algorithmic}
\usepackage{graphicx}
\usepackage{textcomp}
\usepackage{caption}
\usepackage{subcaption}
\usepackage{xspace} 
\usepackage{placeins}
 \def\BibTeX{{\rm B\kern-.05em{\sc i\kern-.025em b}\kern-.08em
     T\kern-.1667em\lower.7ex\hbox{E}\kern-.125emX}}
 \newcommand{\safeincludegraphics}[2]{%
     \IfFileExists{#2}{\includegraphics[width=#1]{#2}}{%
         \fbox{\parbox[c][0.16\textheight][c]{#1}{\centering Missing file\\\texttt{#2}}}%
     }%
 }
\begin{document}
\begin{acronym}
	\acro{IRS}{Intelligent reflecting surface}
	\acro{RIS}{reconfigurable intelligent surface}
	\acro{irs}{intelligent reflecting surface}
	\acro{PARAFAC}{parallel factor}
	\acro{NPF}{Nested PARAFAC2}
	\acro{TALS}{trilinear alternating least squares}
	\acro{DF}{decode-and-forward}
	\acro{AF}{amplify-and-forward}
	\acro{CE}{channel estimation}
	\acro{RF}{radio-frequency}
	\acro{THz}{Terahertz communication}
	\acro{EVD}{eigenvalue decomposition}
	\acro{CRB}{Cramér-Rao lower bound}
	\acro{CSI}{channel state information}
	\acro{BS}{base station}
	\acro{MIMO}{multiple-input multiple-output}
	\acro{NMSE}{normalized mean squared error}
	\acro{2G}{Second Generation}
	\acro{3G}{3$^\text{rd}$~Generation}
	\acro{3GPP}{3$^\text{rd}$~Generation Partnership Project}
	\acro{4G}{4$^\text{th}$~Generation}
	\acro{5G}{5$^\text{th}$~Generation}
	\acro{6G}{6$^\text{th}$~generation}
	\acro{E-TALS}{\textit{enhanced} TALS}
	\acro{UT}{user terminal}
	\acro{UTs}{user terminals}
	\acro{LS}{least squares}
	\acro{KRF}{Khatri-Rao factorization}
	\acro{KF}{Kronecker factorization}
	\acro{MU-MIMO}{multi-user multiple-input multiple-output}
	\acro{MU-MISO}{multi-user multiple-input single-output}
	\acro{MU}{multi-user}
	\acro{SER}{symbol error rate}
	\acro{SNR}{signal-to-noise ratio}
	\acro{SVD}{singular value decomposition}
    \acro{FA}{fluid antenna}
    \acro{FAs}{fluid antennas}
    \acro{PF}{PARAFAC}
    \acro{TTS}{two-time-scale}
    \acro{STS}{single-time-scale}
    
\end{acronym}
%\receiveddate{XX Month, XXXX}
%\reviseddate{XX Month, XXXX}
%\accepteddate{XX Month, XXXX}
%\publisheddate{XX Month, XXXX}
%\currentdate{11 January, 2024}
%\doiinfo{OJCOMS.2024.011100}

% \title{Semi-Blind Receivers for RIS-Aided\\ Fluid Antenna Systems}

% \author{Josué V. de Araújo\IEEEauthorrefmark{1}\IEEEmembership{(Student Member, IEEE)}, Gilderlan T. de Araújo\IEEEauthorrefmark{2}\IEEEmembership{(Member, IEEE)},\\
% André L. F. de Almeida.\IEEEauthorrefmark{1}
% \IEEEmembership{(Senior Member, IEEE)}}
%  \affil{Federal University of Ceará, Fortaleza-CE, Brazil}
%  \affil{Federal Institute of Ceará,Canindé-CE, Brazil}
% \corresp{Josué V. de Araújo (e-mail: josue.vas@alu.ufc.br).}
% \authornote{This work was supported in part by the National Institute of Science and Technology (INCT-Signals) sponsored by Brazil's National Council for Scientific and Technological Development (CNPq) (grants 406517/2022-3, 151870/2025-0 and 303356/2025-1), and FUNCAP (grant 25255-82587.32.41/64).}
% \markboth{Preparation of Papers for IEEE OPEN JOURNALS}{Author \textit{et al.}}

 %\title{Joint Channel and Symbol Estimation for RIS-Empowered Fluid Antenna Systems}
 \title{Semi-Blind Receivers for RIS-Aided\\ Fluid Antenna Systems}

 \author{Josué V. de Araújo, Gilderlan T. de Araújo and André L. F. de Almeida
 %\thanks{This work was supported in part by the National Institute of Science and Technology (INCT-Signals) sponsored by Brazil's National Council for Scientific and Technological Development (CNPq) (projects 406517/2022-3 and 303356/2025-1), and FUNCAP (project ITR-0214-00041.01.00/23).}
 }
 
\maketitle

\begin{abstract}
Reconfigurable intelligent surfaces (RISs) and fluid antennas (FAs) are key technologies for enhancing spatial degrees of freedom in future wireless networks. However, channel acquisition in RIS-aided FA systems is challenging as cascaded links depend on time-varying antenna-port selections and RIS configurations, leading to high training overhead in conventional pilot-based methods. We propose a semi-blind estimation framework for this joint architecture to estimate channels and symbols concurrently. Two hierarchical transmission protocols are introduced, resulting in distinct tensor models. Protocol 1 uses a two-time-scale structure yielding a PARAFAC (PF) model, while Protocol 2 employs a single-time-scale structure with blockwise spatial variations, leading to a Nested PARAFAC2 (NPF) model. For both, we develop semi-blind receivers based on trilinear alternating least squares to jointly estimate user-to-RIS channels, RIS-to-BS channels, and transmitted symbols by exploiting spatio-temporal diversity from FA and RIS reconfiguration. We derive identifiability conditions and computational complexity, revealing a fundamental trade-off: the PF receiver (Protocol 1) more aggressively exploits joint RIS/FA reconfiguration for stronger robustness, whereas the NPF receiver (Protocol 2) offers a flexible, lower-complexity alternative. Simulations show the proposed receivers achieve accurate recovery with significantly reduced training overhead, demonstrating the effectiveness of tensor-based semi-blind processing for RIS-aided fluid antenna communications.
\end{abstract}

\begin{IEEEkeywords}
Fluid antennas, reconfigurable intelligent surface, multiuser systems, tensor decomposition, semi-blind estimation, PARAFAC (PF), Nested PARAFAC2 (NPF).
\end{IEEEkeywords}

%\maketitle

%\renewcommand\baselinestretch{.85}

\section{Introduction}

\IEEEPARstart{W}{ith} the advent of sixth-generation (6G) wireless networks, future transceivers are expected to support extreme operating regimes involving massive connectivity, ultra-high data rates, stringent latency, and high energy efficiency \cite{Saad6G2020,Tataria6G2021}. Meeting these requirements calls for more spectrum and larger arrays, but also for mechanisms such as \ac{RIS}s and \acp{FA} that can actively exploit the spatial degrees of freedom of the propagation channel \cite{Basar2021,Wong_FAMA_2022}. Conventional MIMO and massive MIMO systems already provide substantial multiplexing and beamforming gains, especially in mmWave and THz bands, but they still rely on fixed-position antennas (FPAs), which sample the electromagnetic field at predetermined points and therefore limit the spatial diversity that can be harvested within a finite aperture \cite{historico_movable,Lipeng_2023,New_Tutorial_2025}.

In this context, fluid antennas (FAs) have recently emerged as a promising transceiver architecture \cite{Wong_FAMA_2022,historico_movable,New_Tutorial_2025}. Much of the FA literature has formalized fluid antenna multiple access, characterized the outage and diversity behavior of FA links, and clarified practical operating modes such as slow port adaptation \cite{Wong_FAMA_2022,Wong_sFAMA_2023,New_Wong_2024}. By allowing the radiating element, or equivalently the active RF connection, to be repositioned within a confined region or switched among dense candidate ports, these architectures create an additional spatial adaptation dimension that can be exploited to strengthen desired links and suppress interference \cite{Wong_FAMA_2022,Lipeng_2023}. This repositioning can be realized mechanically, through fluid implementations, or electronically via dense ports and fast RF switching networks \cite{historico_movable,New_Tutorial_2025}. As a result, FA systems can achieve sizeable gains in SINR, coverage, and spectral efficiency without proportionally increasing the number of RF chains \cite{Lipeng_2023, New_Wong_2024}.

In parallel, \ac{RIS}s have become a key enabler of smart radio environments \cite{Basar2021,DiRenzo2020_SmartRadio}. By adaptively controlling the phase, and in some architectures also the amplitude, of the reflected waves, an RIS can reshape the propagation channel, enhance coverage, and create favorable virtual links with very low power consumption \cite{Wu2019_RIS,zheng_survey}. The joint use of FAs and RISs is therefore particularly appealing: while the RIS reconfigures the wireless environment, the FA array reconfigures the effective channel sampling points at the transceiver. Recent works on this joint architecture have mainly focused on beamforming and position optimization, including sum-rate maximization in multiuser settings, joint transmit/passive beamforming and antenna positioning, and analyses of when FAs are beneficial in RIS-assisted links \cite{Sun2025RISMA,Zhang2025RISMA,Wei2025RISMA,Yu2025RISMA}. These studies confirm the potential of RIS-aided FA systems, but they generally assume the availability of channel state information and do not address the estimation problem itself.

By contrast, channel estimation has been extensively investigated for RIS-assisted systems, ranging from pilot-based strategies \cite{Wang2020,Wei2021,Hanzo2022,zheng_survey} to structured tensor models \cite{Gil_SAM,Gil_JTSP,Sokal_BDRIS_2025}. For instance, \cite{Gil_JTSP} formulated a PARAFAC-based framework to decouple the cascaded channels in RIS-MIMO links, while \cite{Sokal_BDRIS_2025} extended tensor modeling to beyond-diagonal RIS architectures. In parallel, channel estimation has also been actively explored for FA systems using compressed sensing \cite{Ma_2023_CE,Xiao_2024}, Bayesian learning \cite{Zhang_2025}, and tensor-based techniques \cite{Ruoyu_2024}. Specifically, \cite{Xiao_2024} exploits the spatial correlation of movable antennas via compressed sensing to estimate the field response, whereas \cite{Ruoyu_2024} models the movable-antenna MIMO channel as a tensor to enable efficient parameter recovery. Furthermore, semi-blind tensor receivers have shown strong performance in RIS-assisted systems without FAs \cite{Gil022,Gil_TSP}, allowing joint channel and symbol estimation with reduced training overhead. Nevertheless, these two research lines remain largely disconnected: existing RIS channel-estimation methods do not account for the observation-dependent channel variations induced by FA reconfiguration, whereas current FA channel-estimation methods do not address RIS-aided cascaded links. Unlike the aforementioned works, semi-blind channel estimation for a joint architecture combining both an RIS and fluid antennas at the transceiver has not yet been addressed, which strongly motivates our tensor-based formulation.

From a signal processing viewpoint, this gap is especially relevant because the joint RIS-FA uplink naturally exhibits a multi-way structure involving antenna ports, RIS configurations, transmission blocks, and symbol times. More importantly, the FA array, while providing an additional design degree of freedom, changes the active receive positions over the observation window, directly reshaping the effective channel seen by the base station and, consequently, the way in which the unknown symbols are embedded in the received data. Tensor decompositions have proven to be attractive to solve joint channel and symbol estimation problems in multi-antenna wireless communications since they capture the underlying multidimensional signal structure, preserving the intrinsic coupling among dimensions and enabling blind or semi-blind recovery of channels and symbols under powerful identifiability properties \cite{Sidiropoulos2017,Andre_overview,DEALMEIDA2007337,4524041,Almeida2013,Favier_TSP_2014,Ximenes_2014,Ximenes2015}. In particular, the PARAFAC2 model \cite{parafac2,Unicidade_PARAFAC2} is well suited to scenarios in which one factor varies across slices, which matches the case of dynamic FA selection across transmission blocks. Although this decomposition remains largely unexplored in wireless communication problems, it has been successfully exploited in MIMO system contexts \cite{sorensen_parafac2_2009,Gil_PARAFAC2_2017}.

Motivated by the above observations, this paper studies the joint channel and symbol estimation problem for an RIS-aided \ac{MU} uplink system equipped with FAs at the \ac{BS}. The central challenge is that the effective uplink channel is jointly shaped by the RIS reflection pattern and by the FA port-selection process, which makes the received data depend on multiple coupled spatial and temporal dimensions while rendering conventional pilot-based estimation increasingly expensive. To address this issue, we propose a semi-blind tensor-based framework that jointly estimates the user-to-RIS channel, the RIS-to-BS channel, and the transmitted symbols by explicitly exploiting the structured diversity induced by RIS reconfiguration and FA mobility. To the best of our knowledge, this is the first work to investigate a joint semi-blind estimation framework for a multiuser uplink communication system that simultaneously combines an RIS and a fluid-antenna array at the base station, and the first to cast this problem into a tensor-based formulation through two complementary signaling protocols and their associated PF- and NPF-based receivers. 

The main contributions are summarized as follows:
\begin{itemize}
    \item We propose two hierarchical temporal transmission protocols for RIS-aided FA systems. The protocols induce different observation structures and expose different trade-offs between signaling overhead, synchronization requirements, and spatial-temporal diversity.
    \item We show that Protocol 1 follows a \ac{TTS} signaling structure and gives rise to a conventional \ac{PF} tensor model, whereas Protocol 2 follows an \ac{STS} signaling structure and yields an \ac{NPF} tensor model. Based on these two decompositions, we refer to the corresponding semi-blind estimators as the \ac{PF} receiver and the \ac{NPF} receiver, respectively.
    \item For both protocols, we develop semi-blind receivers based on trilinear alternating least squares algorithms, which jointly estimate the user-to-RIS channel, the RIS-to-BS channel, and the transmitted symbols without assuming prior knowledge of pilot sequences, while explicitly leveraging the diversity induced by RIS reconfiguration and FA port selection.
    \item We derive identifiability conditions and computational complexity expressions for both protocols, thereby revealing a clear trade-off between the two receivers: the PF receiver associated with Protocol 1 benefits from stronger recoverability and robustness thanks to a more aggressive joint exploitation of RIS/FA reconfiguration, whereas the NPF receiver associated with Protocol 2 provides a lighter and more flexible alternative with reduced implementation burden.
\end{itemize}

The remainder of this paper is organized as follows: Section II provides the tensor preliminaries and reviews the employed decompositions. Section III presents the system model and the two proposed transmission protocols. Section IV develops the tensor signal models associated with the proposed protocols, which serve as a basis for the formulation of our receivers. The tensor-based semi-blind receivers and their estimation steps are derived in Section V. Section VI is dedicated to the analysis of identifiability conditions and computational complexity. Section VII presents the simulation results, and the paper concludes in Section VIII.

\vspace{2ex}
\noindent \textit{Notation and properties}: Vectors are denoted by boldface lowercase letters ($\mathbf{a}$). Matrices are in boldface capital letters ($\mathbf{A}$), while tensors are symbolized by calligraphic letters $(\boldsymbol{\mathcal{A}})$. The transpose and pseudo-inverse of a matrix $\mathbf{A}$ are denoted as $\mathbf{A}^{\text{T}}$ and $\mathbf{A}^\dagger$, respectively. $\mathbf{I}_{N}$ denotes a $N \times N$ identity matrix. $D_i(\mathbf{A})$ is a diagonal matrix that holds the $i$-th row of $\mathbf{A}$ on its main diagonal. The operator $\textrm{diag}(\mathbf{a})$ forms a diagonal matrix from its vector argument. $\|\cdot\|_{\text{F}}$ denotes the Frobenius norm. The symbol $\otimes$ represents the Kronecker product, while $\diamond$ denotes the Khatri-Rao (column-wise Kronecker) product. The operator $\textrm{vec}(\cdot)$ vectorizes an $I \times J$ matrix, while $\textrm{unvec}_{I \times J}(\cdot)$ does the opposite operation. Moreover, $\mathbf{A}_{(i,j)}$, $\mathbf{A}_{(i,:)}$, and $\mathbf{A}_{(:,j)}$ denote the $(i,j)$-th entry, the $i$-th row, and the $j$-th column of the matrix $\mathbf{A}$, respectively. In this paper, we make use of the following identities:
\begin{equation}
\textrm{vec}(\mathbf{A}\mathbf{B}\mathbf{C}) = (\mathbf{C}^{\textrm{T}} \otimes \mathbf{A})\textrm{vec}(\mathbf{B}),
\label{Eq:Propertie_Vec_General}
\end{equation}
\begin{equation}
(\mathbf{A} \otimes \mathbf{B})(\mathbf{C} \otimes \mathbf{D}) = (\mathbf{A}\mathbf{C}) \otimes (\mathbf{B}\mathbf{D}),
\label{Eq:Propertie_Mixed}
\end{equation}
\begin{equation}
\textrm{vec}(\mathbf{A}\textrm{diag}(\mathbf{x})\mathbf{B}) = (\mathbf{B}^{\textrm{T}} \diamond \mathbf{A})\mathbf{x},
\label{Eq:Propertie_KhatriRao}
\end{equation}
where the matrices and vectors involved have compatible dimensions.

\section{Tensor preliminaries}
%In this section, we summarize the fundamental tensor operations and decompositions utilized throughout this work \cite{Kolda_2009,comon_2009,Andre_overview,Chen2021,Sokal_BDRIS_2025}. We adopt the standard notation where the tensors are denoted by calligraphic letters (e.g., $\boldsymbol{\mathcal{X}}$), the matrices are denoted by uppercase boldface letters (e.g., $\mathbf{X}$), and the vectors by lowercase boldface letters (e.g., $\mathbf{x}$).
In this section, we review the main tensor concepts and decompositions used throughout this work, with particular emphasis on the PARAFAC and PARAFAC2 models.

%\subsection{Slices and Unfoldings}
%A third-order tensor $\boldsymbol{\mathcal{X}} \in \mathbb{C}^{I \times J \times K}$ can be partitioned into frontal slices, denoted as $\mathbf{X}_k \in \mathbb{C}^{I \times J}$ for $k=1, \ldots, K$, by fixing the third dimension index. To apply linear algebra tools to multi-way data, the tensor can be reordered into matrices through the unfolding operation in $n$-mode, denoted as $[\boldsymbol{\mathcal{X}}]_{(n)}$ \cite{Kolda_2009,comon_2009}. Specifically, the 1-mode, 2-mode, and 3-mode unfoldings are defined as the concatenation of the frontal slices along different orientations:
%\begin{align}
 %   [\boldsymbol{\mathcal{X}}]_{(1)} &= [\mathbf{X}_1, \ldots, \mathbf{X}_K] \in \mathbb{C}^{I \times JK}, \\
  %  [\boldsymbol{\mathcal{X}}]_{(2)} &= [\mathbf{X}_1^{\text{T}}, \ldots, \mathbf{X}_K^{\text{T}}] \in \mathbb{C}^{J \times IK}, \\
  %  [\boldsymbol{\mathcal{X}}]_{(3)} &= [\text{vec}(\mathbf{X}_1), \ldots, \text{vec}(\mathbf{X}_K)]^{\text{T}} \in \mathbb{C}^{K \times IJ}.
%\end{align}
%Furthermore, the $n$-mode product between a tensor $\boldsymbol{\mathcal{X}}$ and a matrix $\mathbf{A} \in \mathbb{C}^{L \times I_n}$ is denoted by $\boldsymbol{\mathcal{Y}} = \boldsymbol{\mathcal{X}} \times_n \mathbf{A}$, where $I_n$ matches the $n$-th dimension of the tensor. This operation is equivalent to matrix multiplication $[\boldsymbol{\mathcal{Y}}]_{(n)} = \mathbf{A}[\boldsymbol{\mathcal{X}}]_{(n)}$.

\subsection{PARAFAC Decomposition}
The conventional \textit{parallel factor} (PARAFAC) decomposition represents a third-order tensor as the superposition of rank-one components \cite{Harshman70,Kolda_2009}. For a tensor $\boldsymbol{\mathcal{X}} \in \mathbb{C}^{I_1 \times I_2 \times I_3}$ of rank $R$, its frontal slices can be written as
\begin{equation}
    \mathbf{X}_{i_3} = \mathbf{A}D_{i_3}(\mathbf{C}) \mathbf{B}^\text{T}, \,\,\, i_3=1, \ldots, I_3,
    \label{EQ: generic CP model}
\end{equation}
where $\mathbf{A} \in \mathbb{C}^{I_1 \times R}$, $\mathbf{B} \in \mathbb{C}^{I_2 \times R}$, and $\mathbf{C} \in \mathbb{C}^{I_3 \times R}$ are the factor matrices associated with the three tensor modes. 
%Equivalently, the tensor can be expressed as
% \begin{equation}
%     \boldsymbol{\mathcal{X}} = \sum_{r=1}^{R} \mathbf{a}_r \circ \mathbf{b}_r \circ \mathbf{c}_r,
% \end{equation}
% where $\circ$ denotes the outer product, and $\mathbf{a}_r$, $\mathbf{b}_r$, and $\mathbf{c}_r$ are the $r$-th columns of $\mathbf{A}$, $\mathbf{B}$, and $\mathbf{C}$, respectively. 
A key feature of the PARAFAC model is that the factor matrices remain fixed across all slices, which makes it especially suitable for regular tensors. Under mild conditions, this decomposition is essentially unique up to permutation and scaling ambiguities. In particular, Kruskal's condition $k_{\mathbf{A}} + k_{\mathbf{B}} + k_{\mathbf{C}} \geq 2R + 2$ guarantees uniqueness, where $k_{\mathbf{A}}$, $k_{\mathbf{B}}$, and $k_{\mathbf{C}}$ denote the Kruskal ranks of the corresponding factor matrices \cite{sorensen_parafac2_2009}.

\subsection{PARAFAC2 Decomposition}
PARAFAC2 \cite{parafac2} can be seen as a tensor decomposition that relaxes some constraints of the conventional \textit{parallel factor} (PARAFAC) decomposition \cite{Harshman70}. More specifically, whereas in the PARAFAC model two factor matrices remain constant along two dimensions (forming a regular tensor), the PARAFAC2 decomposition allows handling a set of variant matrices (along the first dimension) with the same number of columns but different row sizes \cite{sorensen_parafac2_2009,Gil_PARAFAC2_2017}. To make the understanding more straightforward, the PARAFAC2 decomposition of a third-order tensor can be represented in terms of its frontal slice notation according to the following expression
\begin{equation}
    \mathbf{X}_{i_3} = \mathbf{A}_{i_3}D_{i_3}(\mathbf{C}) \mathbf{B}^\text{T}, \,\,\, i_3=1, \ldots, I_3,
    \label{EQ: generic PARAFAC model}
\end{equation}
where the 3-mode factor matrix is $\mathbf{A}_{i_3} \in \mathbb{C}^{I_{i_3} \times R}$, while $\mathbf{B} \in \mathbb{C}^{I_2 \times R}$ and $\mathbf{C} \in \mathbb{C}^{I_3 \times R}$ are fixed factor matrices. Generally, the PARAFAC2 decomposition presented in (\ref{EQ: generic PARAFAC model}) is not unique \cite{Kolda_2009}. However, under additional constraints, uniqueness can be reached by imposing $\mathbf{A}_{i_3}^{\text{T}}\mathbf{A}_{i_3} = \boldsymbol{\Phi}$, $i_3 = 1, \ldots, I_3$. This means $\mathbf{A}_{i_3} = \mathbf{M}_{i_3}\mathbf{N}$, where $\mathbf{M}_{i_3}^{\text{H}}\mathbf{M}_{i_3} = \mathbf{I}_{I_{i_3}}$ \cite{Unicidade_PARAFAC2}.

\subsection{Nested PARAFAC2 Decomposition}
In some applications, the slice-dependent factor of an outer PARAFAC2 model may itself admit a PARAFAC2 structure. This leads to a nested or hybrid construction in which two coupled PARAFAC2 models appear at different levels of the decomposition. A convenient way to describe this situation is to write the outer model as
\begin{equation}
    \mathbf{X}_{i_3} = \mathbf{A}_{i_3} D_{i_3}(\mathbf{C}) \mathbf{B}^{\text{T}}, \qquad i_3 = 1, \ldots, I_3,
    \label{EQ: nested outer PARAFAC2}
\end{equation}
and then assume that the varying factor $\mathbf{A}_{i_3}$ also follows a PARAFAC2-type factorization, namely
\begin{equation}
    \mathbf{A}_{i_3} = \mathbf{M}_{i_3} D_{i_3}(\boldsymbol{\Gamma}) \mathbf{N}^{\text{T}}, \qquad i_3 = 1, \ldots, I_3,
    \label{EQ: nested inner PARAFAC2}
\end{equation}
where $\mathbf{M}_{i_3}$ is slice-dependent, while $\mathbf{N}$ and $\boldsymbol{\Gamma}$ play the role of common latent factors for the inner structure. By substituting (\ref{EQ: nested inner PARAFAC2}) into (\ref{EQ: nested outer PARAFAC2}), one obtains
\begin{equation}
    \mathbf{X}_{i_3} = \mathbf{M}_{i_3} D_{i_3}(\boldsymbol{\Gamma}) \mathbf{N}^{\text{T}} D_{i_3}(\mathbf{C}) \mathbf{B}^{\text{T}},
    \label{EQ: nested PARAFAC2 model}
\end{equation}
which makes explicit that the overall model is obtained by nesting one PARAFAC2 decomposition inside another. In other words, the outer tensor follows a PARAFAC2 model, but its slice-dependent factor is not free, since it is itself generated by a second PARAFAC2 model. This type of representation is useful when the physical system exhibits two coupled sources of slice variation, as occurs in the RIS-aided fluid-antenna setting studied in this paper.

\section{System Model}
We consider a \ac{MU} uplink communication system where $K$ single-antenna users transmit data to a \ac{BS}  assisted by a \ac{RIS}. The BS is equipped with a one-dimensional fluid-antenna (FA) architecture comprising $N$ available ports, each connected to $M$ antennas via a fast-switching mechanism, with $M \leq N$. Among the $N$ available ports, $M$ are selected to be active and connected to the \ac{RF} chains. This port selection is governed by a binary selection matrix $\mathbf{S} \in \{0,1\}^{M \times N}$. Since only one antenna can be active per port, the matrix structure follows $\mathbf{S}_{(m,n)} \in \{0,1\}$, where a value of $1$ indicates that the $m$-th antenna is connected to the $n$-th port, with the constraint that only one element per column is allowed. Consequently, the selection matrix satisfies $\mathbf{S} \mathbf{S}^{\mathrm{H}} = \mathbf{I}_M$, and the following constraints apply:
\begin{equation}
    \|\mathbf{S}_{(:,m)}\| = \|\mathbf{S}_{(n,:)}\| = 1.
    \label{EQ: S constraint}
\end{equation} 
Based on this architecture, we investigate and analyze the impact of two proposed temporal transmission protocols.
\begin{figure}[!t]
    \centering
    \subfloat[]{%
        \includegraphics[width=0.49\textwidth]{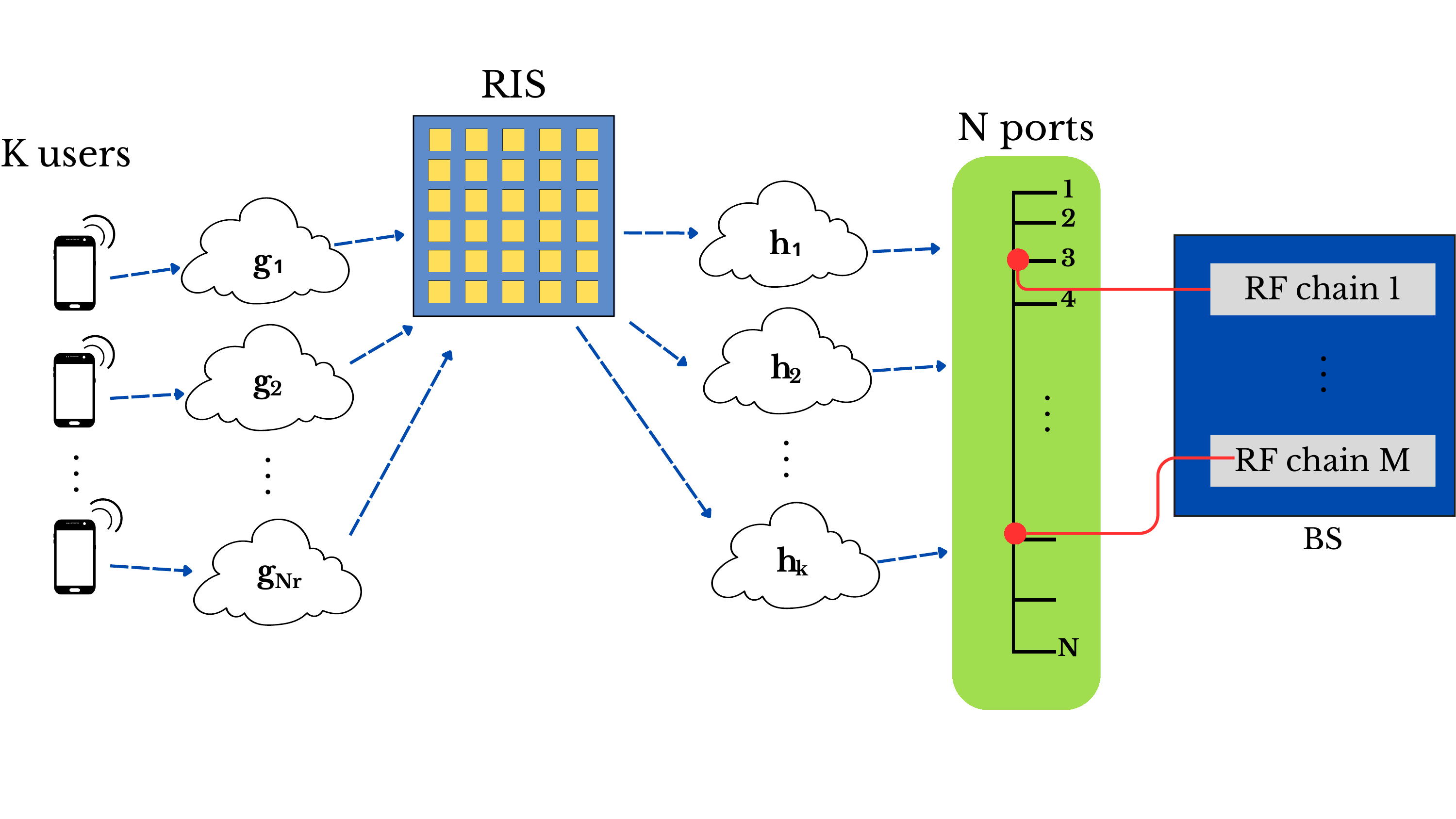}
        \label{fig:system}}
    \hfill
    \subfloat[]{%
        \includegraphics[width=0.49\textwidth]{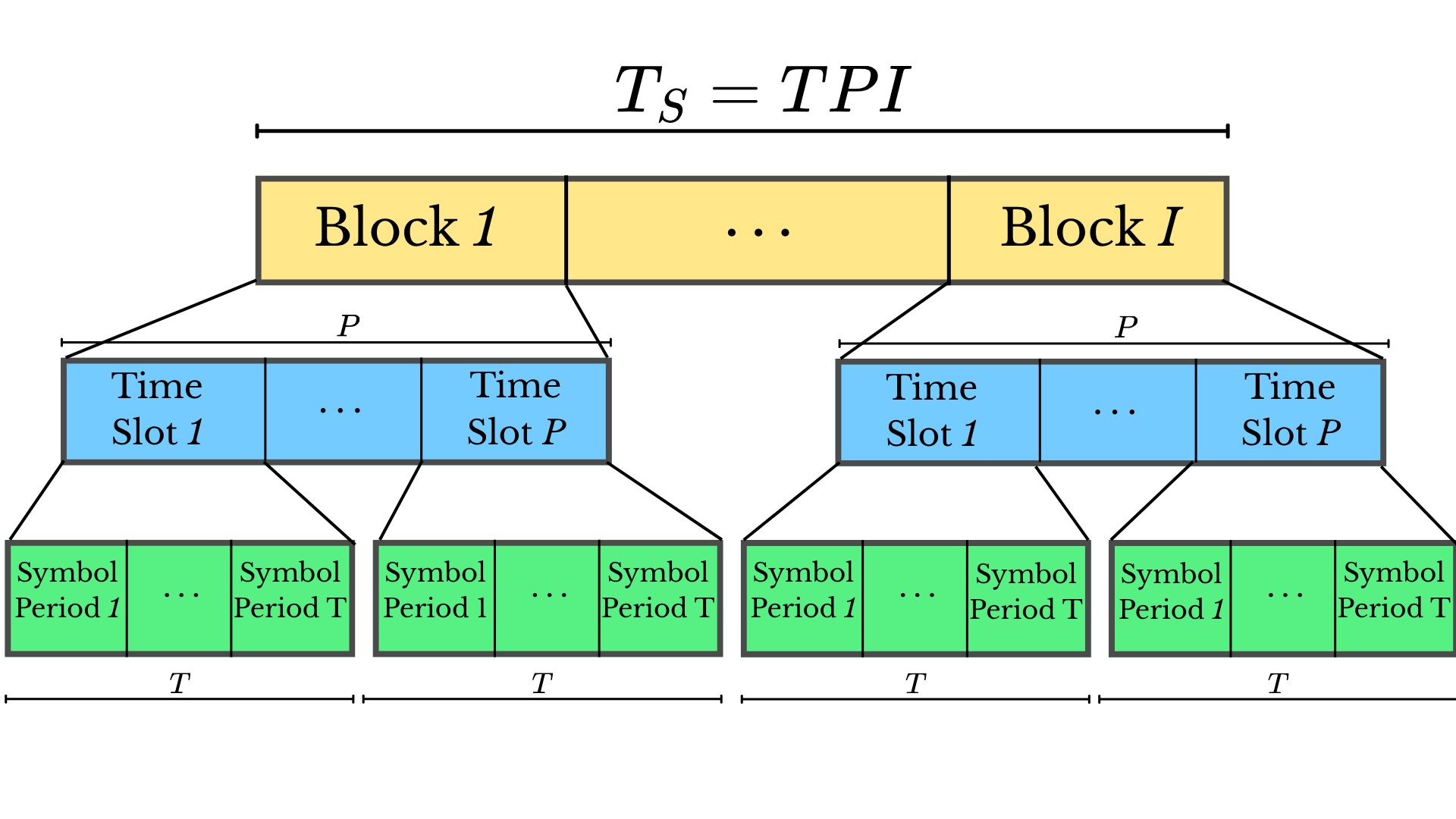}
        \label{fig:time}}
    \caption{System model and transmission structure: (a) fluid-antenna-based uplink \ac{MU} scenario; (b) transmission time structure.}
    \label{fig:geral}
\end{figure}
\subsection{Protocol 1: \ac{TTS} Structure}

In the first protocol, the transmission is organized according to a \ac{TTS} structure composed of $I$ spatial blocks, each consisting of $P$ time slots and $T$ symbol periods. In this scheme, the FA port selection matrix $\mathbf{S}_i \in \{0,1\}^{M \times N}$ and the RIS phase-shifting matrix $D_i(\mathbf{\Theta}) \in \mathbb{C}^{N_r \times N_r}$ remain invariant across all time slots of the block $i$, but vary between different blocks. In contrast, the temporal coding matrix $D_p(\mathbf{C}) \in \mathbb{C}^{K \times K}$ independently scales the symbols at each $p$-th time slot. This allows the BS to capture $I$ distinct spatial observations in each slot, effectively decoupling the spatial diversity of the fluid antennas and the RIS from temporal coding variations. In this setup, the received baseband signal vector $\mathbf{y}_{i,p,t}$ in the $t$-th symbol period is modeled by superposition of signals from all $K$ users reflected by the $N_r$ RIS elements:
\begin{equation}
    \mathbf{y}_{i,p,t} = \mathbf{S}_i \sum_{n_r = 1}^{N_r} \mathbf{h}_{n_r} \theta_{i,n_r} \sum_{k = 1}^{K} g_{n_r,k} x_{k,t} + \mathbf{z}_{i,p,t} \in \mathbb{C}^{M \times 1}, 
\end{equation}
where $\theta_{i,n_r} \in \mathbb{C}$ is the phase-shifting coefficient of the $n_r$-th RIS element for the $i$-th block configuration. By grouping the terms associated with the cascaded channel matrices, the summation can be eliminated to express the signal more compactly in vector form:
\begin{equation}
    \mathbf{y}_{i,p,t} = \mathbf{S}_{i} \mathbf{H} D_{i}(\mathbf{\Theta}) \mathbf{G} \mathbf{x}_{t} + \mathbf{z}_{i,p,t}, %\in \mathbb{C}^{M \times 1},
\end{equation}
where $D_{i}(\mathbf{\Theta}) = \text{diag}(\theta_{i,1}, \dots, \theta_{i,N_r})$ contains the phase-shifting coefficients specifically for the $i$-th block. To incorporate the diversity across time slots, the symbols for each user are encoded across the $P$ time slots, leading to the following temporal-coded model:
\begin{equation}
    \mathbf{y}_{i,p,t} = \mathbf{S}_{i} \mathbf{H} D_{i}(\mathbf{\Theta}) \mathbf{G} D_{p}(\mathbf{C}) \mathbf{x}_{t} + \mathbf{z}_{i,p,t}.% \in \mathbb{C}^{M \times 1}.
\end{equation}
Collecting the received signal vectors over all $T$ symbol periods, we obtain the aggregated matrix representation for the $(i, p)$-th component:
\begin{equation}
    \mathbf{Y}_{i,p} = \mathbf{S}_{i} \mathbf{H} D_{i}(\mathbf{\Theta}) \mathbf{G} D_{p}(\mathbf{C}) \mathbf{X} + \mathbf{Z}_{i,p} \in \mathbb{C}^{M \times T},
    \label{eq9}
\end{equation}
where $\mathbf{Y}_{i,p} \triangleq [\mathbf{y}_{i,p,1}, \dots, \mathbf{y}_{i,p,T}]$ represents the observation matrix for the $i$-th block within the $p$-th time slot. The \ac{MU} data symbol matrix $\mathbf{X} \triangleq [\mathbf{x}_{1}, \dots, \mathbf{x}_{T}] \in \mathbb{C}^{K \times T}$ aggregates the symbols transmitted by all $K$ users, while $\mathbf{Z}_{i,p} \triangleq [\mathbf{z}_{i,p,1}, \dots, \mathbf{z}_{i,p,T}] \in \mathbb{C}^{M \times T}$ denotes the corresponding AWGN matrix. Furthermore, the coding matrix $\mathbf{C} \in \mathbb{C}^{P \times K}$ is defined such that its $p$-th row, $\mathbf{c}_{p,:} = [c_{p,1}, \dots, c_{p,K}]$, contains the set of complex coefficients that scale the users' signals during the $p$-th time slot.

%By vertically stacking the local observations for all $I$ blocks, we define a global multi-component received matrix $\mathbf{Y}_p \triangleq [\mathbf{Y}_{1,p}^T, \dots, \mathbf{Y}_{I,p}^T]^T \in \mathbb{C}^{IM \times T}$ for each time slot $p$. Consequently, the entire set of received signals can be organized into a third-order tensor $\boldsymbol{\mathcal{Y}} \in \mathbb{C}^{IM \times T \times P}$ that strictly follows a canonical PARAFAC model. Unlike the \ac{NPF} structure observed in Protocol 2, the synchronized variation of the RIS phase-shifting values and FA port selections allows the spatial characteristics to be captured by a constant global spatial factor matrix $\mathbf{W} \triangleq [\mathbf{Q}_1^T, \dots, \mathbf{Q}_I^T]^T \mathbf{G} \in \mathbb{C}^{IM \times K}$, where the auxiliary matrix is given by $\mathbf{Q}_i \triangleq \mathbf{S}_i \mathbf{H} D_i(\mathbf{\Theta}) \in \mathbb{C}^{M \times N_r}$. Thus, the following precise correspondences with the canonical PARAFAC decomposition factors are established:
%\begin{equation}
 %   \big(\mathbf{A} \; , \mathbf{B} \; , \mathbf{C}\big) \leftrightarrow \big(\mathbf{W} \; , \mathbf{X}^T \; , \mathbf{C}\big).
%\end{equation}

\subsection{Protocol 2: \ac{STS} Structure}

In the second protocol, an \ac{STS} block-fading channel model is adopted over a transmission frame comprising $I$ transmission blocks, each spanning $T$ symbol periods. During the $i$-th block, a block of $T$ symbols is transmitted while the FA port selection matrix $\mathbf{S}_i \in \{0,1\}^{M \times N}$ is kept constant. However, to fully exploit spatial diversity, this selection matrix is dynamically updated across different blocks. The base-band received signal vector at the BS for the $t$-th symbol period of the $i$-th block is modeled as:
\begin{equation}
    \mathbf{y}_{i,t} = \mathbf{S}_i \sum_{n_r = 1}^{N_r} \mathbf{h}_{n_r} \theta_{i,n_r} \sum_{k = 1}^{K} g_{n_r,k} x_{k,t} + \mathbf{z}_{i,t} \in \mathbb{C}^{M \times 1}, 
    \label{EQ_Received_Signal_SUM}
\end{equation}
where $\mathbf{h}_{n_r} \in \mathbb{C}^{N \times 1}$ denotes the $n_r$-th column of the RIS-to-BS channel matrix $\mathbf{H} \in \mathbb{C}^{N \times N_r}$, $\theta_{i,n_r} \in \mathbb{C}$ is the phase-shifting coefficient of the $n_r$-th RIS element during block $i$, $g_{n_r,k} \in \mathbb{C}$ is the channel gain between the $k$-th user and the $n_r$-th RIS element, and $x_{k,t} \in \mathbb{C}$ is the data symbol transmitted by user $k$ at time $t$.

By defining the user-to-RIS channel matrix as $\mathbf{G} = [\mathbf{g}_{1}, \dots, \mathbf{g}_{K}] \in \mathbb{C}^{N_r \times K}$ and the vector of \ac{MU} transmitted symbol as $\mathbf{x}_{t} = [x_{1,t}, \dots, x_{K,t}]^T \in \mathbb{C}^{K \times 1}$, the received signal can be compactly rewritten in vector form as:
\begin{equation}
    \mathbf{y}_{i,t} = \mathbf{S}_{i} \mathbf{H} D_{i}(\mathbf{\Theta}) \mathbf{G} \mathbf{x}_{t} + \mathbf{z}_{i,t} \in \mathbb{C}^{M \times 1},
\end{equation}
where $D_{i}(\mathbf{\Theta}) = \text{diag}(\theta_{i,1}, \dots, \theta_{i,N_r}) \in \mathbb{C}^{N_r \times N_r}$ contains the reflection coefficients corresponding to the $i$-th row of the global RIS configuration matrix $\mathbf{\Theta} \in \mathbb{C}^{I \times N_r}$. To efficiently exploit multi-block temporal diversity, transmitted symbols are spread across the $I$ blocks using a temporal coding matrix $\mathbf{C} \in \mathbb{C}^{I \times K}$. Incorporating this coding gain for the $i$-th block yields:
\begin{equation}
    \mathbf{y}_{i,t} = \mathbf{S}_{i} \mathbf{H} D_{i}(\mathbf{\Theta}) \mathbf{G} D_{i}(\mathbf{C}) \mathbf{x}_{t} + \mathbf{z}_{i,t}, %\in \mathbb{C}^{M \times 1},
\end{equation}
where $D_i(\mathbf{C}) = \text{diag}(c_{i,1}, \dots, c_{i,K}) \in \mathbb{C}^{K \times K}$ is formed from the $i$-th row of $\mathbf{C}$. Collecting the received signal vectors over all $T$ symbol periods within the $i$-th block, we obtain the aggregated observation matrix $\mathbf{Y}_{i}$:
\begin{equation}
    \mathbf{Y}_{i} = \mathbf{S}_{i} \mathbf{H} D_{i}(\mathbf{\Theta}) \mathbf{G} D_{i}(\mathbf{C}) \mathbf{X} + \mathbf{Z}_{i} \in \mathbb{C}^{M \times T},
    \label{eq8}
\end{equation}
where $\mathbf{Y}_{i} \triangleq [\mathbf{y}_{i,1}, \dots, \mathbf{y}_{i,T}]$ aggregates the spatial measurements during the coherence interval. The \ac{MU} data matrix $\mathbf{X} \triangleq [\mathbf{x}_{1}, \dots, \mathbf{x}_{T}] \in \mathbb{C}^{K \times T}$ collects transmitted symbols, and $\mathbf{Z}_{i} \triangleq [\mathbf{z}_{i,1}, \dots, \mathbf{z}_{i,T}] \in \mathbb{C}^{M \times T}$ denotes the additive white Gaussian noise matrix (AWGN).

%It should be noted that the set of noiseless received signal matrices modeled in \eqref{eq8} can be interpreted as the $p$-th frontal slice ($p = 1, \dots, P$) of a third-order observation tensor $\boldsymbol{\mathcal{Y}} \in \mathbb{C}^{M \times T \times P}$ that strictly follows a PARAFAC2 decomposition model. By analyzing the cascaded channel structure, the precise correspondences between the physical signal model and the mathematical PARAFAC2 factors are established as:
%\begin{equation}
 %   \big(\mathbf{A}_{p} \; , \mathbf{B} \; , \mathbf{C}\big) \leftrightarrow \big(\mathbf{S}_{p}\mathbf{H} D_p(\mathbf{\Theta})\mathbf{G} \; , \mathbf{X}^T \; , \mathbf{C}\big).
%\end{equation}
%In this multilinear representation, the effective spatial factor $\mathbf{A}_p \in \mathbb{C}^{M \times K}$ captures the dynamic physical interaction between the FA selection matrix $\mathbf{S}_p$ and the RIS.parameterized cascade channel. Consequently, the intrinsic coupling between the user-to-RIS gains in $\mathbf{G}$ and the RIS-to-BS channel $\mathbf{H}$ ensures the structural uniqueness required to perform the tensor decomposition and accurately estimate the channel parameters.
\begin{table*}
\centering
\caption{Comparison of signal models, tensor dimensions, and decompositions for the proposed protocols.}
\label{table:comparison_protocols}
\begin{tabular}{|c|c|c|c|c|}
\hline
\textbf{Protocol} & \textbf{Temporal Variation} & \textbf{Tensor Model} & \textbf{Tensor Dimension} & \textbf{Uniqueness Condition} \\ \hline
Protocol 1 & $\mathbf{S}_i, \mathbf{\Theta}_i$ (block $i$), $\mathbf{C}_p$ (time slot $p$) & \begin{tabular}{@{}c@{}} PARAFAC (\ac{PF}) \\ $\boldsymbol{\mathcal{Y}} \sim \Big( \mathbf{W}, \mathbf{X}^T, \mathbf{C} \Big)$ \end{tabular} & $IM \times T \times P$ & $IMTP \geq N_r\max(K, N)$, $IM \geq N_r$ \\ \hline
Protocol 2 & $\mathbf{S}_i, \mathbf{\Theta}_i, \mathbf{C}_i$ block $i$ & \begin{tabular}{@{}c@{}} Nested PARAFAC2 (\ac{NPF}) \\ $\boldsymbol{\mathcal{Y}} \sim \Big( \{ \mathbf{A}_i \}_{i=1}^I, \mathbf{X}^T, \mathbf{C} \Big)$ \end{tabular} & $M \times T \times I$ & $MTI \geq N_r\max( K, N)$, $IM \geq  K$ \\ \hline
\end{tabular}
\end{table*}

\section{Tensor signal modeling}

To accurately model the multidimensional nature of the cascaded RIS-assisted communication channel, the received signal for each proposed transmission protocol is rigorously recast into tensor frameworks. In this section, we explicitly detail the tensor slices, the mode unfoldings (matricizations), and the multilinear factorizations, following standard tensor algebra conventions.

\subsection{\ac{PF} Model (Protocol 1)}
In Protocol 1, the FA port selection $\mathbf{S}_i$ and the RIS phase-shifting values $\mathbf{\Theta}_i$ vary over $I$ blocks, while the user coding $\mathbf{C}_p$ varies over $P$ time slots within each block. We start by defining the noiseless local received signal matrix $\mathbf{Y}_{i,p} \in \mathbb{C}^{M \times T}$ for the $i$-th block and $p$-th coding slot:
\begin{equation}\label{eq:cpprot2}
    \mathbf{Y}_{i,p} = \mathbf{S}_i \mathbf{H} D_i(\mathbf{\Theta}) \mathbf{G} D_p(\mathbf{C}) \mathbf{X} \in \mathbb{C}^{M \times T}.
\end{equation}

For a given coding slot $p$, we vertically stack the local observations across all $i = 1, \dots, I$ blocks to form a global spatial slice $\bar{\mathbf{Y}}_p \in \mathbb{C}^{IM \times T}$:
\begin{equation}
    \bar{\mathbf{Y}}_p = \begin{bmatrix} \mathbf{Y}_{1,p} \\ \vdots \\ \mathbf{Y}_{I,p} \end{bmatrix} = \begin{bmatrix} \mathbf{S}_1 \mathbf{H} D_1(\mathbf{\Theta}) \\ \vdots \\ \mathbf{S}_I \mathbf{H} D_I(\mathbf{\Theta}) \end{bmatrix} \mathbf{G} D_p(\mathbf{C}) \mathbf{X} \in \mathbb{C}^{IM \times T}.
\end{equation}
In this vertical stacking structure, the $M$ spatial dimensions associated with the active FA ports vary the fastest (acting as the inner index within each block), while the block index $i$ varies the slowest (acting as the outer index tracking the stacked blocks).

By defining the constant global spatial factor matrix $\mathbf{W} \in \mathbb{C}^{IM \times K}$ as
\begin{equation}
    \mathbf{W} \triangleq \begin{bmatrix} \mathbf{S}_1 \mathbf{H} D_1(\mathbf{\Theta}) \\ \vdots \\ \mathbf{S}_I \mathbf{H} D_I(\mathbf{\Theta}) \end{bmatrix} \mathbf{G} \in \mathbb{C}^{IM \times K},
\end{equation}
the global spatial slice can be rewritten directly in terms of the temporal symbol matrix $\mathbf{X} \in \mathbb{C}^{K \times T}$ in a compact form:
\begin{equation}
    \bar{\mathbf{Y}}_p = \mathbf{W} D_p(\mathbf{C}) \mathbf{X} \in \mathbb{C}^{IM \times T}.
\end{equation}

This matrix constitutes the $p$-th frontal slice of a third-order tensor $\boldsymbol{\mathcal{Y}} \in \mathbb{C}^{IM \times T \times P}$. Because the global spatial signature $\mathbf{W}$ is invariant with respect to the coding time slot $p$, the tensor $\boldsymbol{\mathcal{Y}}$ strictly follows a \ac{PF} decomposition model.

To perform joint channel and symbol estimation, we exploit the standard mode-3 unfolding property of the PF tensor, $\mathbf{Y}_{(3)} \in \mathbb{C}^{P \times IMT}$, which is mathematically governed by the Khatri-Rao product ($\diamond$):
\begin{equation}
    \mathbf{Y}_{(3)} = \mathbf{C} (\mathbf{X}^T \diamond \mathbf{W})^T \in \mathbb{C}^{P \times IMT}.
\end{equation}
% According to standard tensor algebra conventions, the Khatri--Rao product determines the ordering used in the vectorization. In $(\mathbf{X}^T \diamond \mathbf{W})$, the entries associated with the global spatial factor $\mathbf{W} \in \mathbb{C}^{IM \times K}$ are stacked first, so the spatial index of size $IM$ varies fastest. By contrast, the temporal index associated with $\mathbf{X}^T$ is stacked afterward and therefore varies more slowly.
According to the PARAFAC model introduced in Section II.A, the correspondence between the slice expression $\bar{\mathbf{Y}}_p = \mathbf{W} D_p(\mathbf{C}) \mathbf{X}$ and the generic representation in \eqref{EQ: generic CP model} is given by $\big(\mathbf{X}_{i_3}, \mathbf{A}, \mathbf{B}, \mathbf{C}\big) \leftrightarrow \big(\bar{\mathbf{Y}}_p, \mathbf{W}, \mathbf{X}^{\mathrm{T}}, \mathbf{C}\big)$. Hence, the frontal slice index $i_3$ of the generic PARAFAC model is identified here with the coding-slot index $p$, while the three factor matrices are respectively associated with the global spatial factor $\mathbf{W}$, the symbol factor $\mathbf{X}^{\mathrm{T}}$, and the coding matrix $\mathbf{C}$.

The final representation, mapping the cascaded physical parameters to the associated PARAFAC model is given as:
\begin{equation}
    \boldsymbol{\mathcal{Y}} \sim \text{PF}\Big( \mathbf{W}, \mathbf{X}^T, \mathbf{C} \Big) \in \mathbb{C}^{IM \times T \times P},
\end{equation}
where the global spatial factor $\mathbf{W}$ is explicitly given by:
\begin{equation}
    \mathbf{W} = \begin{bmatrix} \mathbf{S}_1 \mathbf{H} D_1(\mathbf{\Theta}) \\ \vdots \\ \mathbf{S}_I \mathbf{H} D_I(\mathbf{\Theta}) \end{bmatrix} \mathbf{G} \in \mathbb{C}^{IM \times K}.
\end{equation}
This formulation explicitly reveals how the local spatial interactions at the FA array and RIS are deterministically nested within the global PF structure.

\subsection{\ac{NPF} Model (Protocol 2)}
Let us define the noiseless received signal matrix at the $i$-th block (for $i = 1, \dots, I$). Starting from the physical system model, the $i$-th frontal slice $\mathbf{Y}_i \in \mathbb{C}^{M \times T}$ of the observation tensor $\boldsymbol{\mathcal{Y}} \in \mathbb{C}^{M \times T \times I}$ is expressed as:
\begin{equation}
    \mathbf{Y}_i = \mathbf{S}_i \mathbf{H} D_i(\mathbf{\Theta}) \mathbf{G} D_i(\mathbf{C}) \mathbf{X} \in \mathbb{C}^{M \times T},
\end{equation}
where $D_i(\cdot)$ constructs a diagonal matrix from the $i$-th row of its argument. By defining the time-varying effective spatial factor $\mathbf{A}_i \in \mathbb{C}^{M \times K}$ as
\begin{equation}
    \mathbf{A}_i \triangleq \mathbf{S}_i \mathbf{H} D_i(\mathbf{\Theta}) \mathbf{G} \in \mathbb{C}^{M \times K},
\end{equation}
we can rewrite the slice directly in terms of the temporal symbol matrix $\mathbf{X} \in \mathbb{C}^{K \times T}$ as:
\begin{equation}
    \mathbf{Y}_i = \mathbf{A}_i D_i(\mathbf{C}) \mathbf{X} \in \mathbb{C}^{M \times T}.
\end{equation}

Following standard tensor unfolding definitions, by vertically stacking the slices along the block dimension $I$, we obtain the mode-1 matricization of the tensor, denoted as $\mathbf{Y}_{(1)} \in \mathbb{C}^{IM \times T}$:
\begin{equation}
    \mathbf{Y}_{(1)} = \begin{bmatrix} \mathbf{Y}_1 \\ \vdots \\ \mathbf{Y}_I \end{bmatrix} = \begin{bmatrix} \mathbf{A}_1 D_1(\mathbf{C}) \\ \vdots \\ \mathbf{A}_I D_I(\mathbf{C}) \end{bmatrix} \mathbf{X} \in \mathbb{C}^{IM \times T}.
\end{equation}
In this block-wise stacking structure, the spatial dimension associated with the $M$ active FA ports varies the fastest (acting as the inner index within each block $\mathbf{Y}_i$), while the block index $i$ varies the slowest (acting as the outer index tracking the stacked blocks). 

Crucially, the effective spatial signature matrix $\mathbf{A}_i$ is not an unstructured slice-dependent factor. Indeed, the outer relation
\begin{equation}
    \mathbf{Y}_i = \mathbf{A}_i D_i(\mathbf{C}) \mathbf{X}
    \label{EQ: outer_protocol1_parafac2}
\end{equation}
is itself a PARAFAC2 model, while the factor $\mathbf{A}_i$ admits the structured decomposition
\begin{equation}
    \mathbf{A}_i = \mathbf{S}_i \mathbf{H} D_i(\mathbf{\Theta}) \mathbf{G},
    \label{EQ: inner_protocol1_parafac2}
\end{equation}
which also has a PARAFAC2 form because $\mathbf{S}_i$ and $D_i(\mathbf{\Theta})$ vary with the slice index $i$, whereas $\mathbf{H}$ and $\mathbf{G}$ remain common across slices. Therefore, the model in the preceding equation for $\mathbf{Y}_i$ can be interpreted as a nesting of two PARAFAC2 models: an outer PARAFAC2 decomposition for $\boldsymbol{\mathcal{Y}}$, and an inner PARAFAC2 decomposition governing the effective factor $\mathbf{A}_i$.

According to the Nested PARAFAC2 model introduced in Section II.C, the correspondences associated with \eqref{EQ: outer_protocol1_parafac2} and \eqref{EQ: inner_protocol1_parafac2} are given by $\big(\mathbf{X}_{i_3}, \mathbf{A}_{i_3}, \mathbf{B}, \mathbf{C}\big) \leftrightarrow \big(\mathbf{Y}_i, \mathbf{A}_i, \mathbf{X}^{\mathrm{T}}, \mathbf{C}\big)$
% \begin{equation}
%     \big(\mathbf{X}_{i_3}, \mathbf{A}_{i_3}, \mathbf{B}, \mathbf{C}\big) \leftrightarrow \big(\mathbf{Y}_i, \mathbf{A}_i, \mathbf{X}^{\mathrm{T}}, \mathbf{C}\big),
% \end{equation}
for the outer PARAFAC2 model, and $\big(\mathbf{A}_{i_3}, \mathbf{M}_{i_3}, \boldsymbol{\Gamma}, \mathbf{N}\big) \leftrightarrow \big(\mathbf{A}_i, \mathbf{S}_i\mathbf{H}, \mathbf{\Theta}, \mathbf{G}^{\mathrm{T}}\big)$
% \begin{equation}
%     \big(\mathbf{A}_{i_3}, \mathbf{M}_{i_3}, \boldsymbol{\Gamma}, \mathbf{N}\big) \leftrightarrow \big(\mathbf{A}_i, \mathbf{S}_i\mathbf{H}, \mathbf{\Theta}, \mathbf{G}^{\mathrm{T}}\big),
% \end{equation}
for the inner PARAFAC2 model. Hence, the slice-dependent factor of the outer decomposition is itself generated by a second PARAFAC2 structure.

The final representation mapping the physical cascaded parameters to the nested PARAFAC2 model is given as:
\begin{equation}
    \boldsymbol{\mathcal{Y}} \sim \text{NPF}\Big( \big\{ \mathbf{A}_i \big\}_{i=1}^I, \mathbf{X}^T, \mathbf{C} \Big) \in \mathbb{C}^{M \times T \times I},
\end{equation}
where the time-varying effective spatial factor $\mathbf{A}_i$ is physically modeled as:
\begin{equation}
    \mathbf{A}_i = \mathbf{S}_i \mathbf{H} D_i(\mathbf{\Theta}) \mathbf{G} \in \mathbb{C}^{M \times K}.
\end{equation}
% This formulation clearly reveals that the slice dependence induced by $\mathbf{S}_i$ and $\mathbf{\Theta}_i$ generates the inner PARAFAC2 structure, while the coding matrix $\mathbf{C}$ and symbol matrix $\mathbf{X}$ define the outer PARAFAC2 structure.

With the tensor signal models for Protocols 1 and 2 now established, the next section exploits these \ac{PF} and \ac{NPF} structures to derive the corresponding semi-blind receivers.

\section{Tensor-based Semi-Blind Receivers}
For the joint semi-blind estimation of the channels and symbols, we solve the following alternating optimization problems, where the Frobenius norm yields a scalar cost function in $\mathbb{R}$:
\begin{equation}
    \Big(\hat{\mathbf{H}}, \hat{\mathbf{G}}, \; \hat{\mathbf{X}}\Big)  = \underset{\mathbf{H},\mathbf{G}, \mathbf{X}}{\arg\min} \,\, \left\|\mathbf{Y}_p - \mathbf{S}_p \mathbf{H} \mathbf{D}_p(\mathbf{\Theta})\mathbf{G}\mathbf{D}_p(\mathbf{C}) \mathbf{X}\right\|_\text{F}^2.
    \label{Eq:argmin1}
\end{equation}
\begin{equation}
    \Big(\hat{\mathbf{H}}, \hat{\mathbf{G}}, \; \hat{\mathbf{X}}\Big)  = \underset{\mathbf{H}, \mathbf{G}, \mathbf{X}}{\arg\min} \,\, \left\|\mathbf{Y}_{i,p} - \mathbf{S}_i \mathbf{H} \mathbf{D}_i(\mathbf{\Theta})\mathbf{G}\mathbf{D}_p(\mathbf{C}) \mathbf{X}\right\|_\text{F}^2.
    \label{Eq:argmin2}
\end{equation}
For notation consistency across both protocols, we use $\mathbf{Y}_{stacked}$ to denote the vertically stacked observation matrix and $\mathbf{B}_{total}$ to denote the corresponding global effective sensing matrix in the symbol-estimation subproblems.
\subsection{\ac{PF} Receiver for Protocol 1}

Recalling the hierarchical frame structure of Protocol 1, the FA port selection matrix $\mathbf{S}_i \in \{0,1\}^{M \times N}$ and the RIS phase-shifting matrix $D_i(\mathbf{\Theta}) \in \mathbb{C}^{N_r \times N_r}$ are fixed during the $i$-th block. Conversely, the temporal coding matrix $D_p(\mathbf{C}) \in \mathbb{C}^{K \times K}$ varies dynamically at each $p$-th time slot. Therefore, the baseband received signal for the $(i, p)$-th component is formulated as:
\begin{equation}
    \mathbf{Y}_{i,p} = \mathbf{S}_i \mathbf{H} D_i(\mathbf{\Theta}) \mathbf{G} D_p(\mathbf{C}) \mathbf{X} + \mathbf{Z}_{i,p} \in \mathbb{C}^{M \times T},
\end{equation}
where $i = 1, \dots, I$ and $p = 1, \dots, P$ denote the spatial block and temporal slot indices, respectively. The matrices $\mathbf{H} \in \mathbb{C}^{N \times N_r}$ and $\mathbf{G} \in \mathbb{C}^{N_r \times K}$ represent the cascaded channels, $\mathbf{X} \in \mathbb{C}^{K \times T}$ collects the transmitted symbols, and $\mathbf{Z}_{i,p} \in \mathbb{C}^{M \times T}$ accounts for the additive white Gaussian noise (AWGN).
By vertically stacking the spatial measurements across all $I$ blocks for a given time slot $p$, we define the aggregated observation matrix $\mathbf{Y}_p$:
\begin{equation}
    \mathbf{Y}_p = 
    \begin{bmatrix} 
        \mathbf{Y}_{1,p} \\ 
        \vdots \\ 
        \mathbf{Y}_{I,p} 
    \end{bmatrix} = 
    \mathbf{W} D_p(\mathbf{C}) \mathbf{X} + \mathbf{Z}_p \in \mathbb{C}^{IM \times T},
\end{equation}
where $\mathbf{Z}_p \in \mathbb{C}^{IM \times T}$ is the corresponding stacked noise matrix. The global spatial factor matrix $\mathbf{W} \in \mathbb{C}^{IM \times K}$ is factorized as $\mathbf{W} \triangleq \mathbf{Q} \mathbf{G}$, where the auxiliary matrix $\mathbf{Q} \in \mathbb{C}^{IM \times N_r}$ encapsulates the aggregated RIS-to-BS channel interactions across all $I$ components, defined as:
\begin{equation}
    \mathbf{Q} \triangleq 
    \begin{bmatrix} 
        \mathbf{S}_1 \mathbf{H} D_1(\mathbf{\Theta}) \\ 
        \vdots \\ 
        \mathbf{S}_I \mathbf{H} D_I(\mathbf{\Theta}) 
    \end{bmatrix} \in \mathbb{C}^{IM \times N_r}.
\end{equation}

Because the spatial factor $\mathbf{W}$ remains invariant with respect to the temporal slot index $p$, the multi-dimensional structure of the aggregated signals naturally conforms to a \ac{PF} decomposition. By collecting the matrices $\mathbf{Y}_p$ along a third mode, we construct the third-order received signal tensor $\boldsymbol{\mathcal{Y}} \in \mathbb{C}^{IM \times T \times P}$. Utilizing the $n$-mode product notation, the tensor model is compactly formulated as:
\begin{equation}
    \boldsymbol{\mathcal{Y}} = \boldsymbol{\mathcal{I}}_{3,K} \times_1 \mathbf{W} \times_2 \mathbf{X}^T \times_3 \mathbf{C},
\end{equation}
where $\boldsymbol{\mathcal{I}}_{3,K}$ denotes the $K \times K \times K$ identity tensor, and $\mathbf{C} \in \mathbb{C}^{P \times K}$ is the temporal coding factor matrix, whose $p$-th row collects the diagonal elements of $D_p(\mathbf{C})$.

% Under Protocol 2, the received data follow the \ac{NPF}-structured model in \eqref{eq8}. Based on this model, the next subsections derive the TALS updates for $\mathbf{G}$, $\mathbf{H}$, and $\mathbf{X}$.

\subsubsection{Estimation of $\mathbf{G}$}
With the \ac{PF} formulation established above, we now derive the TALS update equations for Protocol 1.
To estimate the user-to-RIS channel matrix $\mathbf{G} \in \mathbb{C}^{N_r \times K}$, we reconsider the aggregated received signal model for the $p$-th time slot:
\begin{equation}
    \mathbf{Y}_p = \mathbf{Q} \mathbf{G} D_p(\mathbf{C}) \mathbf{X} + \mathbf{Z}_p \in \mathbb{C}^{IM \times T}.
\end{equation}
To isolate $\mathbf{G}$, we apply the vectorization property defined in \eqref{Eq:Propertie_Vec_General} to obtain the vectorized observation vector $\mathbf{y}_p \triangleq \text{vec}(\mathbf{Y}_p)$:
\begin{equation}
    \mathbf{y}_p = \left( (D_p(\mathbf{C})\mathbf{X})^T \otimes \mathbf{Q} \right) \text{vec}(\mathbf{G}) + \mathbf{z}_p \in \mathbb{C}^{IMT \times 1},
\end{equation}
where $\mathbf{z}_p = \text{vec}(\mathbf{Z}_p) \in \mathbb{C}^{IMT \times 1}$ is the noise vector. Let $\mathbf{g} \triangleq \text{vec}(\mathbf{G}) \in \mathbb{C}^{N_r K \times 1}$. By vertically stacking the observations $\mathbf{y}_1, \dots, \mathbf{y}_P$ across all $P$ time slots, we construct the global linear system for Protocol 1:
\begin{equation}
    \mathbf{y} = 
    \begin{bmatrix}
        \mathbf{y}_1 \\
        \vdots \\
        \mathbf{y}_P
    \end{bmatrix} = 
    \begin{bmatrix}
        (D_1(\mathbf{C})\mathbf{X})^T \otimes \mathbf{Q} \\
        \vdots \\
        (D_P(\mathbf{C})\mathbf{X})^T \otimes \mathbf{Q}
    \end{bmatrix} \mathbf{g} + 
    \begin{bmatrix}
        \mathbf{z}_1 \\
        \vdots \\
        \mathbf{z}_I
    \end{bmatrix} \in \mathbb{C}^{IMTP \times 1}.
\end{equation}
This global system can be compactly rewritten as:
\begin{equation}
    \mathbf{y} = \mathbf{W}_{\mathbf{G}} \mathbf{g} + \mathbf{z} \in \mathbb{C}^{IMTP \times 1},
\end{equation}
where $\mathbf{W}_{\mathbf{G}} \in \mathbb{C}^{IMTP \times N_r K}$ denotes the aggregate regression matrix, and $\mathbf{z} \in \mathbb{C}^{IMTP \times 1}$ is the stacked noise vector. The least squares (LS) estimate for $\mathbf{g}$ is formulated as the following optimization problem:
\begin{equation}
    \hat{\mathbf{g}} = \arg \min_{\mathbf{g}} \left\| \mathbf{y} - \mathbf{W}_{\mathbf{G}} \mathbf{g} \right\|_2^2,
\end{equation}
which is directly solved via the Moore-Penrose pseudo-inverse:
\begin{equation}
    \hat{\mathbf{g}} = \mathbf{W}_{\mathbf{G}}^{\dagger} \mathbf{y}, \quad \text{and} \quad \hat{\mathbf{G}} = \text{unvec}_{N_r \times K}(\hat{\mathbf{g}}).
\end{equation}
\subsubsection{Estimation of $\mathbf{H}$}
To estimate the \ac{BS}-to-\ac{RIS} channel matrix $\mathbf{H} \in \mathbb{C}^{N \times N_r}$, we must isolate it from the spatial and temporal coupling matrices. First, we define the cascaded transmitted signal associated with the $p$-th time slot as:
\begin{equation}
    \mathbf{R}_p \triangleq \mathbf{G} D_p(\mathbf{C}) \mathbf{X} \in \mathbb{C}^{N_r \times T}.
\end{equation}
By concatenating these components across all $P$ time slots, we construct the aggregate temporal matrix $\mathbf{R}_{aux}$:
\begin{equation}
    \mathbf{R}_{aux} \triangleq \begin{bmatrix} \mathbf{R}_1, \dots, \mathbf{R}_P \end{bmatrix} \in \mathbb{C}^{N_r \times PT}.
\end{equation}

By fixing the spatial block index $i$ and aggregating the observations over the $P$ slots, the received signal matrix $\mathbf{Y}_i$ can be rewritten as:
\begin{equation}
    \mathbf{Y}_i = \mathbf{S}_i \mathbf{H} D_i(\mathbf{\Theta}) \mathbf{R}_{aux} + \mathbf{Z}_i \in \mathbb{C}^{M \times PT},
\end{equation}
where $\mathbf{Z}_i \in \mathbb{C}^{M \times PT}$ is the noise matrix corresponding to the $i$-th spatial block. Let $\mathbf{B}_i$ denote the effective incident signal impinging upon the \ac{RIS} during the $i$-th block, defined as:
\begin{equation}
    \mathbf{B}_i \triangleq D_i(\mathbf{\Theta}) \mathbf{R}_{aux} \in \mathbb{C}^{N_r \times PT}.
\end{equation}

Applying the vectorization property to $\mathbf{Y}_i$, we obtain the vectorized observation vector $\mathbf{y}_{\mathbf{H},i}$:
\begin{equation}
    \mathbf{y}_{\mathbf{H},i} \triangleq \text{vec}(\mathbf{Y}_i) = \left( \mathbf{B}_i^T \otimes \mathbf{S}_i \right) \text{vec}(\mathbf{H}) + \mathbf{z}_{\mathbf{H},i} \in \mathbb{C}^{MTP \times 1}.
\end{equation}

Let $\mathbf{h} \triangleq \text{vec}(\mathbf{H}) \in \mathbb{C}^{N N_r \times 1}$, and define the regression sub-matrix for the $i$-th block as $\mathbf{\Psi}_i$:
\begin{equation}
    \mathbf{\Psi}_i \triangleq \mathbf{B}_i^T \otimes \mathbf{S}_i \in \mathbb{C}^{MTP \times N N_r}.
\end{equation}

To fully exploit the spatial diversity provided by the \ac{FA} movement and \ac{RIS} phase variations, we vertically stack the vectorized observations $\mathbf{y}_{\mathbf{H},1}, \dots, \mathbf{y}_{\mathbf{H},I}$ across all $I$ blocks:
\begin{equation}
    \mathbf{y}_{\mathbf{H}} = 
    \begin{bmatrix}
        \mathbf{y}_{\mathbf{H},1} \\
        \vdots \\
        \mathbf{y}_{\mathbf{H},I}
    \end{bmatrix} = 
    \begin{bmatrix}
        \mathbf{\Psi}_1 \\
        \vdots \\
        \mathbf{\Psi}_I
    \end{bmatrix} \mathbf{h} + 
    \begin{bmatrix}
        \mathbf{z}_{\mathbf{H},1} \\
        \vdots \\
        \mathbf{z}_{\mathbf{H},I}
    \end{bmatrix} \in \mathbb{C}^{IMTP \times 1}.
\end{equation}

\begin{algorithm}[!t]
\caption{Semi-blind PF Receiver for Protocol 1}
\label{alg:TALS_P2}
\KwIn{$\mathbf{Y}_{i,p}$, $D_i(\mathbf{\Theta})$, $D_p(\mathbf{C})$, and $\mathbf{S}_i$ for $i=1, \ldots, I$, $p=1, \ldots, P$. Initialize $\hat{\mathbf{X}}$ and $\hat{\mathbf{H}}$ randomly. Set $j=0$ and $\epsilon(0) = \infty$.}
\KwOut{$\hat{\mathbf{G}}$, $\hat{\mathbf{H}}$, $\hat{\mathbf{X}}$}
\Begin{
    \While{$\|\epsilon(j) - \epsilon(j-1)\| \geq \delta$}{
        1. Construct the aggregate regression matrix $\mathbf{W}_{\mathbf{G}}$ using the current estimates $\hat{\mathbf{H}}$ and $\hat{\mathbf{X}}$, then update $\mathbf{g}$:
        \begin{equation*}
            \hat{\mathbf{g}} = \mathbf{W}_{\mathbf{G}}^{\dagger} \mathbf{y},
        \end{equation*}
        \quad and reconstruct $\hat{\mathbf{G}} = \text{unvec}_{N_r \times K}(\hat{\mathbf{g}})$\;
        
        2. Construct the aggregate regression matrix $\mathbf{W}_{\mathbf{H}}$ using the current estimates $\hat{\mathbf{G}}$ and $\hat{\mathbf{X}}$, then update $\mathbf{h}$:
        \begin{equation*}
            \hat{\mathbf{h}} = \mathbf{W}_{\mathbf{H}}^{\dagger} \mathbf{y},
        \end{equation*}
        \quad and reconstruct $\hat{\mathbf{H}} = \text{unvec}_{N \times N_r}(\hat{\mathbf{h}})$\;
        
        3. Construct the global spatial sensing matrix $\mathbf{B}_{total}$ using $\hat{\mathbf{G}}$ and $\hat{\mathbf{H}}$, then update $\mathbf{X}$:
        \begin{equation*}
            \hat{\mathbf{X}} = \mathbf{B}_{total}^{\dagger} \mathbf{Y}_{stacked};
        \end{equation*}
        
        4. Update the iteration counter $j \leftarrow j + 1$\;
        
        5. Compute the data fitting error $\epsilon(j)$:
        \begin{equation*}
            \epsilon(j) = \frac{\|\mathbf{Y}_{stacked} - \mathbf{B}_{total} \hat{\mathbf{X}}\|_F^2}{\|\mathbf{Y}_{stacked}\|_F^2}\;
        \end{equation*}
    }
}
\end{algorithm}

This system is compactly formulated as the global linear model:
\begin{equation}
    \mathbf{y}_{\mathbf{H}} = \mathbf{W}_{\mathbf{H}} \mathbf{h} + \mathbf{z}_{\mathbf{H}} \in \mathbb{C}^{IMTP \times 1},
\end{equation}
where $\mathbf{W}_{\mathbf{H}} \in \mathbb{C}^{IMTP \times N N_r}$ is the comprehensive spatial regression matrix, and $\mathbf{z}_{\mathbf{H}} \in \mathbb{C}^{IMTP \times 1}$ is the stacked noise vector. The least squares (LS) estimation of $\mathbf{h}$ is then formulated by minimizing the residual error:
\begin{equation}
    \hat{\mathbf{h}} = \arg \min_{\mathbf{h}} \left\| \mathbf{y}_{\mathbf{H}} - \mathbf{W}_{\mathbf{H}} \mathbf{h} \right\|_2^2.
\end{equation}

Similarly to the estimation of $\mathbf{G}$, the optimal solution is obtained via the Moore-Penrose pseudo-inverse:
\begin{equation}
    \hat{\mathbf{h}} = \mathbf{W}_{\mathbf{H}}^{\dagger} \mathbf{y}_{\mathbf{H}}, \quad \text{and} \quad \hat{\mathbf{H}} = \text{unvec}_{N \times N_r}(\hat{\mathbf{h}}).
\end{equation}

\subsubsection{Symbol decoding}
To estimate the \ac{MU} symbol matrix $\mathbf{X} \in \mathbb{C}^{K \times T}$, we revisit the aggregated signal model for the $p$-th time slot:
\begin{equation}
    \mathbf{Y}_p = \mathbf{Q} \mathbf{G} D_p(\mathbf{C}) \mathbf{X} + \mathbf{Z}_p \in \mathbb{C}^{IM \times T}.
\end{equation}
Similar to the first protocol, the symbol matrix $\mathbf{X}$ acts as a common right-multiplying factor across all $P$ time slots, inherently bypassing the need for vectorization. We define the effective spatial sensing matrix for the $p$-th slot as $\mathbf{B}_p \triangleq \mathbf{Q} \mathbf{G} D_p(\mathbf{C}) \in \mathbb{C}^{IM \times K}$. By vertically stacking the observation matrices $\mathbf{Y}_1, \dots, \mathbf{Y}_P$, we formulate the global linear system for data estimation:
\begin{equation}
    \mathbf{Y}_{stacked} = 
    \begin{bmatrix}
        \mathbf{Y}_1 \\
        \vdots \\
        \mathbf{Y}_I
    \end{bmatrix} = 
    \begin{bmatrix}
        \mathbf{B}_1 \\
        \vdots \\
        \mathbf{B}_P
    \end{bmatrix} \mathbf{X} + 
    \begin{bmatrix}
        \mathbf{Z}_1 \\
        \vdots \\
        \mathbf{Z}_P
    \end{bmatrix} \in \mathbb{C}^{IMP \times T}.
\end{equation}
This stacked structure can be compactly represented as a standard multiple-input multiple-output (MIMO) model:
\begin{equation}
    \mathbf{Y}_{stacked} = \mathbf{B}_{total} \mathbf{X} + \mathbf{Z}_{noise} \in \mathbb{C}^{IMP \times T},
\end{equation}
where $\mathbf{B}_{total} \in \mathbb{C}^{IMP \times K}$ denotes the global regression matrix and $\mathbf{Z}_{noise} \in \mathbb{C}^{IMP \times T}$ represents the stacked AWGN matrix. The optimization problem to recover the transmitted symbols is formulated as the following least squares (LS) problem based on the Frobenius norm:
\begin{equation}
    \hat{\mathbf{X}} = \underset{\mathbf{X}}{\arg\min} \left\| \mathbf{Y}_{stacked} - \mathbf{B}_{total} \mathbf{X} \right\|_F^2,
\end{equation}
which is directly solved via the Moore-Penrose pseudo-inverse:
\begin{equation}
    \hat{\mathbf{X}} = \mathbf{B}_{total}^{\dagger} \mathbf{Y}_{stacked}.
\end{equation}

With the individual LS estimates $\hat{\mathbf{g}}$, $\hat{\mathbf{h}}$, and $\hat{\mathbf{X}}$ mathematically established, the iterative TALS procedure alternately updates each factor matrix to minimize the global data fitting error while keeping the remaining variables fixed. The complete iterative procedure for Protocol 1 is summarized in \textbf{Algorithm 1}.

\subsection{\ac{NPF} receiver for Protocol 2}
Under Protocol 2, the received data follow the \ac{NPF}-structured model in \eqref{eq8}. Based on this model, the next subsections derive the TALS updates for $\mathbf{G}$, $\mathbf{H}$, and $\mathbf{X}$.

\subsubsection{Estimation of $\mathbf{G}$}
To estimate the channel matrix $\mathbf{G} \in \mathbb{C}^{N_r \times K}$, we start from the received signal model for the $i$-th block:
\begin{equation}
    \mathbf{Y}_{i} = \mathbf{S}_{i} \mathbf{H} D_{i}(\mathbf{\Theta}) \mathbf{G} D_{i}(\mathbf{C}) \mathbf{X} + \mathbf{Z}_{i} \in \mathbb{C}^{M \times T}.
\end{equation}
First, we isolate the term containing $\mathbf{G}$. By applying the vectorization property defined in \eqref{Eq:Propertie_Vec_General}, and considering the diagonal structure of the matrices $D_i(\mathbf{\Theta})$ and $D_i(\mathbf{C})$, we define the vectorized received signal $\mathbf{y}_i = \text{vec}(\mathbf{Y}_i)$:
\begin{equation}
    \mathbf{y}_i = \left( (D_i(\mathbf{C})\mathbf{X})^T \otimes \mathbf{S}_i \mathbf{H} \right) \text{vec}(D_i(\mathbf{\Theta}) \mathbf{G}) + \mathbf{z}_i \in \mathbb{C}^{MT \times 1},
\end{equation}
where $\mathbf{z}_i = \text{vec}(\mathbf{Z}_i) \in \mathbb{C}^{MT \times 1}$. Next, we expand the term $\text{vec}(D_i(\mathbf{\Theta}) \mathbf{G})$ by applying \eqref{Eq:Propertie_Vec_General} once again with an identity matrix $\mathbf{I}_K$, which yields $\text{vec}(D_i(\mathbf{\Theta}) \mathbf{G} \mathbf{I}_K) = (\mathbf{I}_K \otimes D_i(\mathbf{\Theta})) \text{vec}(\mathbf{G})$. Substituting this result back into the previous equation gives:
\begin{equation}
    \mathbf{y}_i = \left( \mathbf{X}^T D_i(\mathbf{C}) \otimes \mathbf{S}_i \mathbf{H} \right) \left( \mathbf{I}_K \otimes D_i(\mathbf{\Theta}) \right) \text{vec}(\mathbf{G}) + \mathbf{z}_i, % \in \mathbb{C}^{MT \times 1}.
\end{equation}
By exploiting the mixed-product property of the Kronecker product given in \eqref{Eq:Propertie_Mixed}, we can further simplify the expression by combining the terms algebraically:
\begin{equation}
    \mathbf{y}_i = \left( \mathbf{X}^T \otimes \mathbf{S}_i \mathbf{H} \right) \left( D_i(\mathbf{C}) \otimes D_i(\mathbf{\Theta}) \right) \text{vec}(\mathbf{G}) + \mathbf{z}_i.% \in \mathbb{C}^{MT \times 1}.
\end{equation}
Let $\mathbf{g} \triangleq \text{vec}(\mathbf{G}) \in \mathbb{C}^{N_r K \times 1}$. By stacking the local observations $\mathbf{y}_1, \dots, \mathbf{y}_I$ vertically, we obtain the complete linear system:
\begin{equation}
    \mathbf{y} = 
    \begin{bmatrix}
        \mathbf{y}_1 \\
        \vdots \\
        \mathbf{y}_I
    \end{bmatrix} = 
    \begin{bmatrix}
        (\mathbf{X}^T \otimes \mathbf{S}_1 \mathbf{H}) (D_1(\mathbf{C}) \otimes D_1(\mathbf{\Theta})) \\
        \vdots \\
        (\mathbf{X}^T \otimes \mathbf{S}_I \mathbf{H}) (D_I(\mathbf{C}) \otimes D_I(\mathbf{\Theta}))
    \end{bmatrix} \mathbf{g} + 
    \begin{bmatrix}
        \mathbf{z}_1 \\
        \vdots \\
        \mathbf{z}_I
    \end{bmatrix} 
\end{equation}
This global system can be compactly written as:
\begin{equation}
    \mathbf{y} = \mathbf{W}_{\mathbf{G}} \mathbf{g} + \mathbf{z} \in \mathbb{C}^{MTI \times 1},
\end{equation}
where $\mathbf{W}_{\mathbf{G}} \in \mathbb{C}^{MTI \times N_r K}$ acts as the global regression matrix, and $\mathbf{z} \in \mathbb{C}^{MTI \times 1}$ represents the stacked noise vectors. The least squares (LS) estimate for $\mathbf{g}$ is obtained by solving the following optimization problem:
\begin{equation}
    \hat{\mathbf{g}} = \arg \min_{\mathbf{g}} \left\| \mathbf{y} - \mathbf{W}_{\mathbf{G}} \mathbf{g} \right\|_2^2,
\end{equation}
which is computed via the Moore-Penrose pseudo-inverse as:
\begin{equation}
    \hat{\mathbf{g}} = \mathbf{W}_{\mathbf{G}}^{\dagger} \mathbf{y}, \quad \text{and} \quad \hat{\mathbf{G}} = \text{unvec}_{N_r \times K}(\hat{\mathbf{g}}).
\end{equation}
\subsubsection{Estimation of $\mathbf{H}$}
To estimate the RIS-to-BS channel matrix $\mathbf{H} \in \mathbb{C}^{N \times N_r}$, we start by recalling the received signal model for the $i$-th block:
\begin{equation}
    \mathbf{Y}_i = \mathbf{S}_i \mathbf{H} D_i(\mathbf{\Theta}) \mathbf{G} D_i(\mathbf{C}) \mathbf{X} + \mathbf{Z}_i \in \mathbb{C}^{M \times T}.
\end{equation}
Our goal is to isolate $\mathbf{H}$. To this end, we group the known parameters on the right side of the channel matrix into an auxiliary matrix $\mathbf{R}_i$, defined as:
\begin{equation}
    \mathbf{R}_i \triangleq D_i(\mathbf{\Theta}) \mathbf{G} D_i(\mathbf{C}) \mathbf{X} \in \mathbb{C}^{N_r \times T}.
\end{equation}
Substituting $\mathbf{R}_i$ back into the signal model yields a simplified expression:
\begin{equation}
    \mathbf{Y}_i = \mathbf{S}_i \mathbf{H} \mathbf{R}_i + \mathbf{Z}_i \in \mathbb{C}^{M \times T}.
\end{equation}
By applying the vectorization property given in \eqref{Eq:Propertie_Vec_General}, we vectorize the observation matrix to obtain $\mathbf{y}_i = \text{vec}(\mathbf{Y}_i)$:
\begin{equation}
    \mathbf{y}_i = \left( \mathbf{R}_i^T \otimes \mathbf{S}_i \right) \text{vec}(\mathbf{H}) + \mathbf{z}_i \in \mathbb{C}^{MT \times 1},
\end{equation}
where $\mathbf{z}_i = \text{vec}(\mathbf{Z}_i) \in \mathbb{C}^{MT \times 1}$ is the noise vector. Let $\mathbf{h} \triangleq \text{vec}(\mathbf{H}) \in \mathbb{C}^{N N_r \times 1}$. To fully exploit the temporal diversity provided by the coding matrix across all $I$ blocks, we stack the local observations $\mathbf{y}_1, \dots, \mathbf{y}_I$ into a global vector $\mathbf{y}$, establishing the complete linear system:
\begin{equation}
    \mathbf{y} = 
    \begin{bmatrix}
        \mathbf{y}_1 \\
        \vdots \\
        \mathbf{y}_I
    \end{bmatrix} = 
    \begin{bmatrix}
        \mathbf{R}_1^T \otimes \mathbf{S}_1 \\
        \vdots \\
        \mathbf{R}_I^T \otimes \mathbf{S}_I
    \end{bmatrix} \mathbf{h} + 
    \begin{bmatrix}
        \mathbf{z}_1 \\
        \vdots \\
        \mathbf{z}_I
    \end{bmatrix} \in \mathbb{C}^{MTI \times 1}.
\end{equation}
This global system can be compactly rewritten as:
\begin{equation}
    \mathbf{y} = \mathbf{W}_{\mathbf{H}} \mathbf{h} + \mathbf{z} \in \mathbb{C}^{MTI \times 1},
\end{equation}
where $\mathbf{W}_{\mathbf{H}} \in \mathbb{C}^{MTI \times N N_r}$ acts as the aggregate regression matrix and $\mathbf{z} \in \mathbb{C}^{MTI \times 1}$ is the global noise vector. The least squares (LS) estimate for the cascaded channel vector $\mathbf{h}$ is found by solving the following optimization problem:
\begin{equation}
    \hat{\mathbf{h}} = \underset{\mathbf{h}}{\arg\min} \left\| \mathbf{y} - \mathbf{W}_{\mathbf{H}}\mathbf{h} \right\|_2^2,
\end{equation}
the solution of which is computed via the Moore-Penrose pseudo-inverse:
\begin{equation}
    \hat{\mathbf{h}} = \mathbf{W}_{\mathbf{H}}^{\dagger} \mathbf{y}, \quad \text{and} \quad \hat{\mathbf{H}} = \text{unvec}_{N \times N_r}(\hat{\mathbf{h}}).
\end{equation}
\subsection{Symbol decoding}
To estimate the \ac{MU} symbol matrix $\mathbf{X} \in \mathbb{C}^{K \times T}$, we revisit the received signal model for the $i$-th block:
\begin{equation}
    \mathbf{Y}_i = \mathbf{S}_i \mathbf{H} D_i(\mathbf{\Theta}) \mathbf{G} D_i(\mathbf{C}) \mathbf{X} + \mathbf{Z}_i \in \mathbb{C}^{M \times T}.
\end{equation}
\begin{algorithm}[!t]
\caption{Semi-blind NPF Receiver for Protocol~2}
\label{alg:TALS_P1}
\KwIn{$\mathbf{Y}_i$, $D_i(\mathbf{\Theta})$, $D_i(\mathbf{C})$, and $\mathbf{S}_i$ for $i=1, \ldots, I$. Initialize $\hat{\mathbf{X}}$ and $\hat{\mathbf{H}}$ randomly. Set $j=0$ and $\epsilon(0) = \infty$.}
\KwOut{$\hat{\mathbf{G}}$, $\hat{\mathbf{H}}$, $\hat{\mathbf{X}}$}
\Begin{
    \While{$\|\epsilon(j) - \epsilon(j-1)\| \geq \delta$}{
        1. Construct the regression matrix $\mathbf{W}_{\mathbf{G}}$ using the current estimates $\hat{\mathbf{H}}$ and $\hat{\mathbf{X}}$, then update $\mathbf{g}$:
        \begin{equation*}
            \hat{\mathbf{g}} = \mathbf{W}_{\mathbf{G}}^{\dagger} \mathbf{y},
        \end{equation*}
        \quad and reconstruct $\hat{\mathbf{G}} = \text{unvec}_{N_r \times K}(\hat{\mathbf{g}})$\;
        
        2. Construct the regression matrix $\mathbf{W}_{\mathbf{H}}$ using the current estimates $\hat{\mathbf{G}}$ and $\hat{\mathbf{X}}$, then update $\mathbf{h}$:
        \begin{equation*}
            \hat{\mathbf{h}} = \mathbf{W}_{\mathbf{H}}^{\dagger} \mathbf{y},
        \end{equation*}
        \quad and reconstruct $\hat{\mathbf{H}} = \text{unvec}_{N \times N_r}(\hat{\mathbf{h}})$\;
        
        3. Construct the global effective sensing matrix $\mathbf{B}_{total}$ using $\hat{\mathbf{G}}$ and $\hat{\mathbf{H}}$, then update $\mathbf{X}$:
        \begin{equation*}
            \hat{\mathbf{X}} = \mathbf{B}_{total}^{\dagger} \mathbf{Y}_{stacked};
        \end{equation*}
        
        4. Update the iteration counter $j \leftarrow j + 1$\;
        
        5. Compute the data fitting error $\epsilon(j)$:
        \begin{equation*}
            \epsilon(j) = \frac{\|\mathbf{Y}_{stacked} - \mathbf{B}_{total} \hat{\mathbf{X}}\|_F^2}{\|\mathbf{Y}_{stacked}\|_F^2}\;
        \end{equation*}
    }
}
\end{algorithm}
Unlike the cascaded channel matrices, the symbol matrix $\mathbf{X}$ does not require vectorization via Kronecker properties, as it inherently acts as a common right-multiplying factor for the known channel parameters across all $I$ blocks. By vertically stacking the observation matrices $\mathbf{Y}_i$, we construct the global linear system directly:
\begin{equation}
    \mathbf{Y}_{stacked} = 
    \begin{bmatrix}
        \mathbf{Y}_1 \\
        \vdots \\
        \mathbf{Y}_I
    \end{bmatrix} = 
    \begin{bmatrix}
        \mathbf{S}_1 \mathbf{H} D_1(\mathbf{\Theta}) \mathbf{G} D_1(\mathbf{C}) \\
        \vdots \\
        \mathbf{S}_I \mathbf{H} D_I(\mathbf{\Theta}) \mathbf{G} D_I(\mathbf{C})
    \end{bmatrix} \mathbf{X} + 
    \begin{bmatrix}
        \mathbf{Z}_1 \\
        \vdots \\
        \mathbf{Z}_I
    \end{bmatrix} 
\end{equation}
This stacked structure can be compactly written as a standard multiple-input multiple-output (MIMO) model:
\begin{equation}
    \mathbf{Y}_{stacked} = \mathbf{B}_{total} \mathbf{X} + \mathbf{Z}_{noise} \in \mathbb{C}^{IM \times T},
\end{equation}
where $\mathbf{B}_{total} \in \mathbb{C}^{IM \times K}$ denotes the global effective sensing matrix, and $\mathbf{Z}_{noise} \in \mathbb{C}^{IM \times T}$ represents the stacked AWGN matrix. The optimization problem for recovering the transmitted symbols is formulated as the following least squares (LS) problem based on the Frobenius norm:
\begin{equation}
    \hat{\mathbf{X}} = \underset{\mathbf{X}}{\arg\min} \left\| \mathbf{Y}_{stacked} - \mathbf{B}_{total} \mathbf{X} \right\|_F^2,
\end{equation}
which is straightforwardly solved via the Moore-Penrose pseudo-inverse:
\begin{equation}
    \hat{\mathbf{X}} = \mathbf{B}_{total}^{\dagger} \mathbf{Y}_{stacked}.
\end{equation}

With the individual LS estimates $\hat{\mathbf{G}}$, $\hat{\mathbf{H}}$, and $\hat{\mathbf{X}}$ established, the iterative TALS procedure alternately updates each factor matrix to minimize the global data fitting error while keeping the remaining variables fixed. The pseudocode for the proposed solution is summarized in \textbf{Algorithm 2}.

\section{Identifiability and Complexity}
In the proposed receivers, identifiability is naturally connected to the least-squares subproblems derived in Section~V, namely the updates of $\mathbf{G}$, $\mathbf{H}$, and $\mathbf{X}$. More precisely, a unique LS update at a given ALS iteration requires the associated regression matrix to be full column rank. In the following, we discuss practical recoverability conditions for the alternating estimation procedure used in the proposed semni-blind receivers. %As in conventional PARAFAC/PARAFAC2-based receivers, the overall factorization remains subject to the usual scaling and permutation ambiguities, which are discussed after the simulation results. 
Before stating compact conditions for each protocol, it is thus convenient to examine the rank requirements induced by the individual LS subproblems.

For Protocol 1, the update of $\mathbf{G}$ relies on $\mathbf{W}_{\mathbf{G}} \in \mathbb{C}^{IMTP \times N_r K}$. A unique LS solution for $\hat{\mathbf{g}}=\mathbf{W}_{\mathbf{G}}^{\dagger}\mathbf{y}$ requires $\mathbf{W}_{\mathbf{G}}$ to have full column rank, i.e., $\mathrm{rank}(\mathbf{W}_{\mathbf{G}})=N_rK$. Since the block rows of $\mathbf{W}_{\mathbf{G}}$ are constructed from matrices of the form $((D_p(\mathbf{C})\mathbf{X})^T \otimes \mathbf{Q})$, this full-rank condition is promoted when the effective spatial factor $\mathbf{Q}$ satisfies $\mathrm{rank}(\mathbf{Q})=N_r$ and the stacked coded-symbol factor built from $\{D_p(\mathbf{C})\mathbf{X}\}_{p=1}^P$ spans the $K$ user dimensions. A necessary dimensional condition is therefore
\begin{equation}
    IMTP \geq N_r K.
\end{equation}
Similarly, since the update of $\mathbf{H}$ is based on $\mathbf{W}_{\mathbf{H}} \in \mathbb{C}^{IMTP \times N N_r}$, a unique LS update requires $\mathrm{rank}(\mathbf{W}_{\mathbf{H}})=NN_r$. Because $\mathbf{W}_{\mathbf{H}}$ is assembled from blocks $\mathbf{B}_i^T \otimes \mathbf{S}_i$, this requirement is met when the effective incident matrices $\{\mathbf{B}_i\}_{i=1}^I$ collectively span the $N_r$-dimensional RIS subspace and the stacked selection operators provide sufficient diversity over the $N$ receive dimensions. This in turn implies the necessary condition
\begin{equation}
    IMTP \geq N N_r.
\end{equation}
For the symbol update, $\mathbf{B}_{total} \in \mathbb{C}^{IMP \times K}$, and uniqueness of $\hat{\mathbf{X}}=\mathbf{B}_{total}^{\dagger}\mathbf{Y}_{stacked}$ requires $\mathrm{rank}(\mathbf{B}_{total})=K$. In other words, the stacked effective channel observed across the $P$ slots must preserve the $K$ user dimensions after RIS coding and FA selection. Hence, a necessary condition is
\begin{equation}
    IMP \geq K.
\end{equation}
Moreover, since the effective spatial factor is built from the stacked RIS/FA observations, a sufficiently rich spatial dimension is also needed to resolve the RIS-related factor, which is summarized by the condition $IM \geq N_r$. Combining the first two inequalities, the channel-estimation stage of Protocol 1 satisfies the global necessary requirement
\begin{equation}
    IMTP \geq \max(N_r K, N N_r) = N_r\max(K,N),
\end{equation}
while $IMP \geq K$ and $IM \geq N_r$ complete the set of practical recoverability conditions. In other words, when the coding matrices, RIS configurations, and FA selection patterns are sufficiently diverse so that $\mathbf{Q}$, the effective incident matrices, and the global sensing matrix $\mathbf{B}_{total}$ attain their target ranks, the corresponding regression matrices become generically full column rank and each LS subproblem admits a unique update. These conditions are summarized in Table I and highlight the main advantage of Protocol 1, which comes from its two-time-scale structure that provides increased diversity, thereby improving the conditioning of the factor updates without increasing the number of unknowns.

For Protocol 2, the update of $\mathbf{G}$ is based on $\mathbf{W}_{\mathbf{G}} \in \mathbb{C}^{MTI \times N_r K}$. A unique LS solution for $\hat{\mathbf{g}} = \mathbf{W}_{\mathbf{G}}^{\dagger}\mathbf{y}$ requires $\mathrm{rank}(\mathbf{W}_{\mathbf{G}})=N_rK$. Since the block rows are built from $((D_i(\mathbf{C})\otimes D_i(\mathbf{\Theta}))^T \otimes (\mathbf{S}_i\mathbf{H}))$, this rank condition is favored when the effective spatial factors $\{\mathbf{S}_i\mathbf{H}\}_{i=1}^I$ span the RIS dimension and the joint coding/RIS coefficients remain sufficiently diverse across blocks. This implies the necessary condition
\begin{equation}
    MTI \geq N_r K.
\end{equation}
Likewise, the update of $\mathbf{H}$ uses $\mathbf{W}_{\mathbf{H}} \in \mathbb{C}^{MTI \times N N_r}$. A unique update requires $\mathrm{rank}(\mathbf{W}_{\mathbf{H}})=NN_r$. Because $\mathbf{W}_{\mathbf{H}}$ is formed from blocks $\mathbf{R}_i^T \otimes \mathbf{S}_i$, this condition is linked to the collective rank of the effective matrices $\mathbf{R}_i$ and to the diversity induced by the selection matrices $\mathbf{S}_i$, which together yield the necessary condition
\begin{equation}
    MTI \geq N N_r.
\end{equation}
Finally, the symbol update is governed by $\mathbf{B}_{total} \in \mathbb{C}^{IM \times K}$, and uniqueness of $\hat{\mathbf{X}} = \mathbf{B}_{total}^{\dagger}\mathbf{Y}_{stacked}$ requires $\mathrm{rank}(\mathbf{B}_{total})=K$, meaning that the stacked effective channel over the $I$ blocks must preserve the $K$ user dimensions. This leads to the necessary condition
\begin{equation}
    IM \geq K.
\end{equation}
Combining the first two inequalities, the channel-estimation stage of Protocol 2 yields the compact necessary condition
\begin{equation}
    MTI \geq \max(N_r K, N N_r) = N_r\max(K,N),
\end{equation}
which, together with $IM \geq K$, gives the practical recoverability requirement for this protocol. As with Protocol 1, these inequalities should be understood as necessary size conditions that support generic full-column-rank behavior of the regression matrices. More specifically, the joint recoverability of channels and symbols is enhanced when the effective spatial channel $\mathbf{S}_i\mathbf{H}$, the incident matrices $\mathbf{R}_i$, and the global sensing matrix $\mathbf{B}_{total}$ attain their target ranks across blocks.

The per-iteration complexity of both algorithms is dominated by the least-squares pseudo-inverse computations associated with the updates of $\mathbf{X}$, $\mathbf{G}$, and $\mathbf{H}$. Using the standard cost of solving an overdetermined LS problem with an $m \times n$ regression matrix, namely $\mathcal{O}(mn^2)$ for $m \geq n$, and neglecting lower-order costs associated with explicitly assembling the structured regression matrices, the complexity expressions in (89) and (90) follow directly from the dimensions of the LS subproblems derived in Section~V. For Protocol 1, the update of $\mathbf{X}$ uses $\mathbf{B}_{total} \in \mathbb{C}^{IMP \times K}$ and therefore costs $\mathcal{O}(IMP K^2)$, while the updates of $\mathbf{G}$ and $\mathbf{H}$ involve $\mathbf{W}_{\mathbf{G}} \in \mathbb{C}^{IMTP \times N_rK}$ and $\mathbf{W}_{\mathbf{H}} \in \mathbb{C}^{IMTP \times NN_r}$, yielding the dominant terms $\mathcal{O}(IMTP N_r^2K^2)$ and $\mathcal{O}(IMTP N^2N_r^2)$, respectively. Hence,
\begin{equation}
    \mathcal{O}_{\text{P1}} = \mathcal{O}\Big(IMP K^2 + IMTP(N_r^2 K^2 + N^2 N_r^2)\Big).
\end{equation}
For Protocol 2, the symbol update uses $\mathbf{B}_{total} \in \mathbb{C}^{IM \times K}$, which gives $\mathcal{O}(IM K^2)$, whereas the channel updates use $\mathbf{W}_{\mathbf{G}} \in \mathbb{C}^{MTI \times N_rK}$ and $\mathbf{W}_{\mathbf{H}} \in \mathbb{C}^{MTI \times NN_r}$, producing the terms $\mathcal{O}(MTI N_r^2K^2)$ and $\mathcal{O}(MTI N^2N_r^2)$. Therefore,
\begin{equation}
    \mathcal{O}_{\text{P2}} = \mathcal{O}\Big(IM K^2 + MTI(N_r^2 K^2 + N^2 N_r^2)\Big).
\end{equation}
If $J$ denotes the number of TALS iterations required for convergence, the total complexity scales as $\mathcal{O}(J\mathcal{O}_{\text{P1}})$ and $\mathcal{O}(J\mathcal{O}_{\text{P2}})$, respectively. Hence, Protocol 2 is computationally lighter, whereas Protocol 1 trades a linear increase with $P$ in the channel-update regressions for improved diversity and, typically, more robust estimation.

\textit{Discussion}: Overall, Protocol~1, associated with the PF receiver, offers the most favorable recoverability conditions because its two-time-scale structure increases the number of observations entering the LS updates of the channel factors, while also relying on a structurally simpler tensor model. From the viewpoint of the RIS-aided FA architecture considered in this paper, this means that the receiver can more effectively exploit the joint diversity created by RIS phase reconfiguration and by the time-varying FA port-selection patterns, which typically results in better conditioning and stronger estimation robustness. This benefit, however, comes at the expense of higher per-iteration complexity and stricter signaling and synchronization requirements, since the RIS coefficients and FA configurations must be coordinated across the two time scales. By contrast, Protocol~2, associated with the NPF receiver, yields a more flexible and computationally lighter solution, since its LS updates involve smaller regression matrices and a single-time-scale organization. In this case, the RIS and FA variations are incorporated into a less demanding signaling structure, which is attractive for practical implementations with tighter overhead or hardware constraints. However, this lower complexity is accompanied by less favorable recoverability margins than those of Protocol~1, so the NPF receiver should be interpreted as a practical low-complexity alternative when implementation simplicity and reduced overhead are prioritized over the stronger identifiability support and robustness enabled by a more aggressive joint exploitation of RIS and FA reconfiguration.

\section{Simulation Results}
\label{sec:results}
\begin{figure*}
    \centering
    \subfloat[]{\safeincludegraphics{0.48\textwidth}{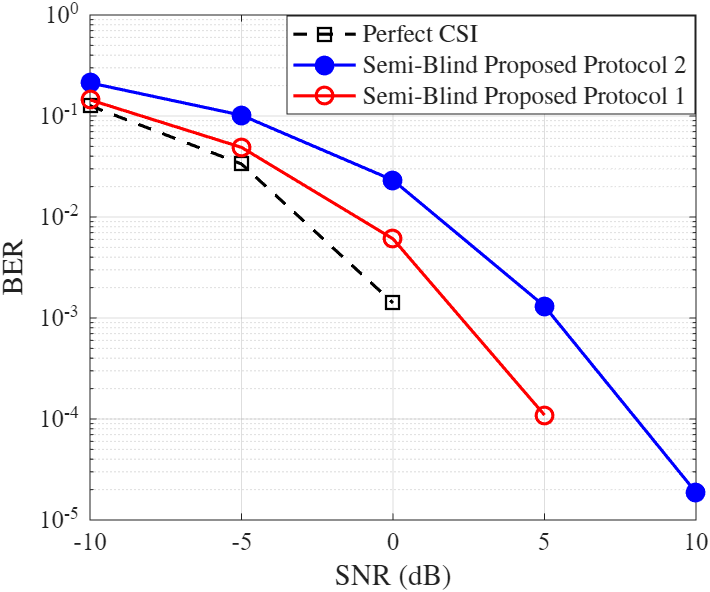}\label{fig:ber_comp}}
    \subfloat[]{\safeincludegraphics{0.48\textwidth}{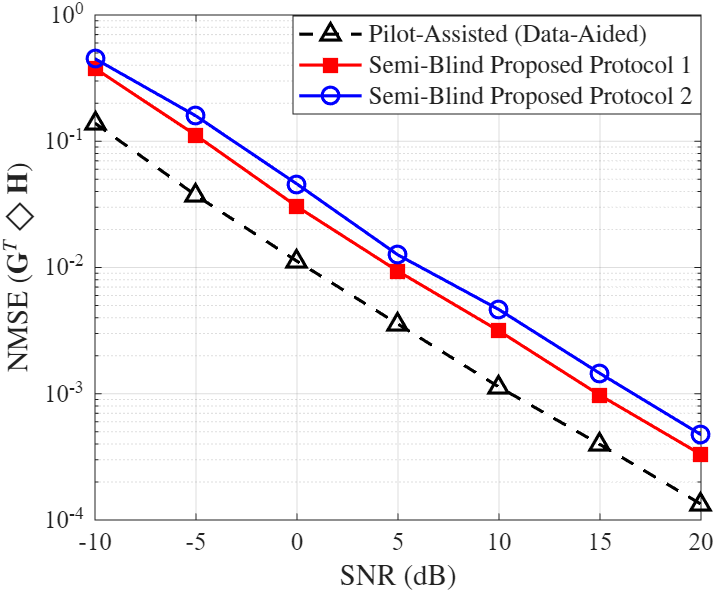}\label{fig:nmse_comp}}
    \caption{Performance comparison of the proposed semi-blind receivers versus SNR: (a) BER for Protocols 1 and 2 with the perfect-CSI benchmark; (b) NMSE of the effective cascaded channel with the pilot-assisted benchmark.}
     \label{fig:comparison}
\end{figure*}

As illustrated in Fig. \ref{fig:comparison}(a), Protocol 1 exhibits a superior symbol recovery capability compared to Protocol 2. The BER curve associated with Protocol 1 converges more rapidly towards lower error floors, closely approaching the theoretical Perfect CSI bound, especially in the mid-to-high SNR regime. A similar trend is corroborated by the channel estimation accuracy depicted in Fig. \ref{fig:comparison}(b), where Protocol 1 yields a consistently lower NMSE than Protocol 2, pushing the performance closer to the ideal Pilot-Assisted (PA) lower bound. It is worth noting that the NMSE is computed over the effective cascaded channel, defined as $\mathbf{H}_{\text{eff}} = \mathbf{G}^T \diamond \mathbf{H}$. Consistently with the identifiability discussion in Section~VI, the individual scaling ambiguities inherent to the alternating least squares (ALS) estimation of $\mathbf{G}$ and $\mathbf{H}$ are aggregated into a single global scaling factor. This overall scaling ambiguity can be effectively resolved at the base station using a minimal number of pilot symbols from $\mathbf{X}$. % thereby validating the practical efficiency of the proposed semi-blind tensor-based estimation.

This performance gain in Protocol 1 is fundamentally attributed to its TTS structure. By varying the RIS phase-shifting matrices and the fluid antenna selection patterns across $I$ blocks, Protocol 1 spans a richer spatial diversity dimension. This enhances the conditioning of the tensor unfolding matrices used to estimate the involved channel and symbol matrices with fewer temporal snapshots, leading to a more robust joint estimation of these quantities. An important practical trade-off emerges from this comparison. The superior performance of Protocol 1 comes at the cost of higher hardware complexity and stricter synchronization requirements, as it necessitates a precise temporal alignment between the variation of the spatial parameters ($I$) and the coding matrices ($P$). Furthermore, the overhead in Protocol 1 scales more aggressively due to its multidimensional block structure. Therefore, while Protocol 1 provides the best overall accuracy, Protocol 2 remains a lower-complexity alternative. In particular, the STS model of Protocol 2 is particularly attractive for practical scenarios where expanding the temporal overhead ($T$) is more feasible than managing the multi-scale signaling and strict synchronization demanded by the TTS architecture.

\begin{figure*}[!t]
    \centering
    \subfloat[]{\safeincludegraphics{0.48\textwidth}{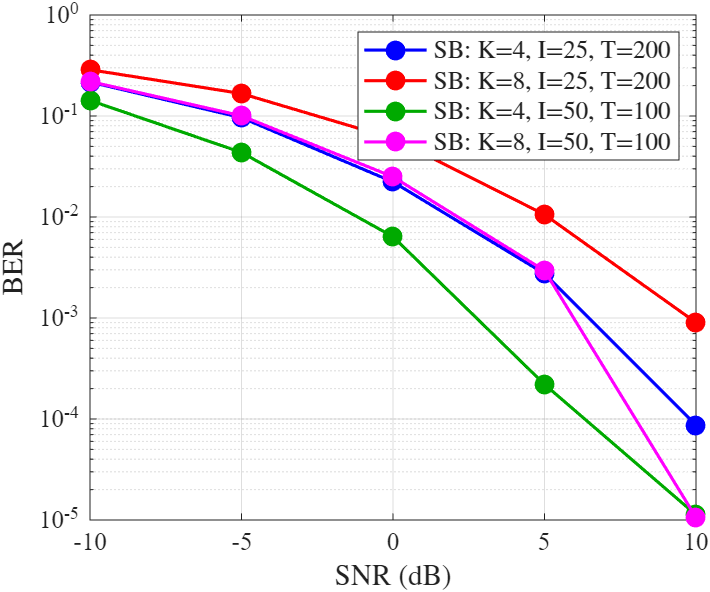}\label{fig:ber_p1_ik}}
    \subfloat[]{\safeincludegraphics{0.48\textwidth}{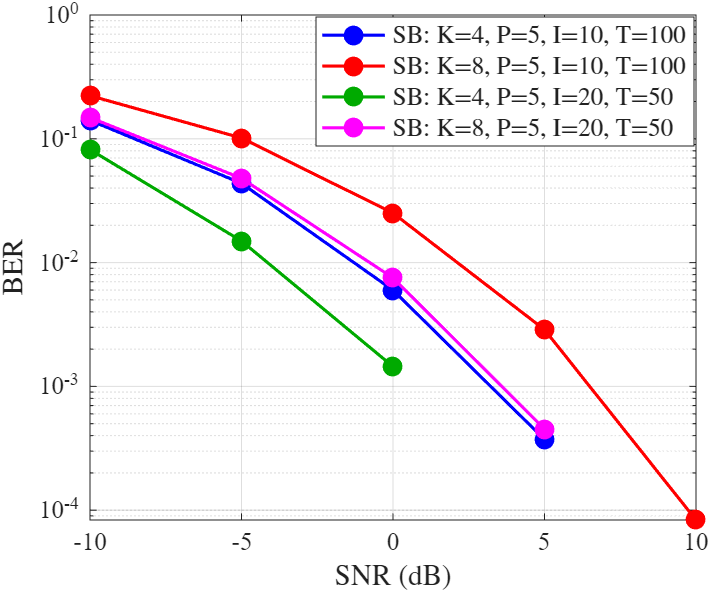}\label{fig:ber_p2_ik}}
    \caption{BER sensitivity to user load and spatial-temporal resource allocation under fixed total overhead: (a) Protocol 2 for different $(K,P,T)$ settings; (b) Protocol 1 for different $(K,I,T)$ settings with $P=5$.}
    \label{fig:ber_ik_results}
\end{figure*}

To further evaluate the robustness of the proposed tensor-based receivers, Fig. \ref{fig:ber_ik_results} investigates the sensitivity of the bit error rate (BER) to variations in the spatial-temporal configuration parameters under a fixed total overhead. Specifically, we evaluate the impact of varying the number of users $K$, the symbol period $T$, and the number of spatial diversity blocks (denoted generically as $P$ or $I$, depending on the protocol). Consistent with the trends previously observed in the NMSE analysis, increasing the number of users from $K=4$ (blue and green curves) to $K=8$ (red and magenta curves) inherently degrades the BER performance across both protocols. This degradation is an expected consequence of the intensified multi-user interference and the increased number of channel coefficients that must be jointly estimated.

An important system design insight can be drawn from the dynamic allocation of the overhead. The results reveal that the performance drop caused by doubling the number of users can be fully compensated by increasing the spatial diversity of the system. For both Protocol 1 (TTS) and Protocol 2 (STS), reallocating the resources by doubling the number of spatial variation blocks (e.g., from 25 to 50) while halving the symbol period $T$ (from 200 to 100) yields a substantial performance improvement. Remarkably, the enhanced spatial degrees of freedom obtained through more frequent updates of the FA selection patterns and RIS phase-shifting values influence the receiver performance. For instance, the configuration serving $K=8$ users with higher spatial diversity (magenta curve) achieves a BER performance comparable to, or even slightly better than, the baseline $K=4$ scenario with longer temporal blocks (blue curve). This consistent behavior across both protocols confirms that maximizing the number of spatial snapshots is more advantageous for reliable symbol recovery than simply extending the number of symbol periods $T$ of each time slot.

\begin{figure*}[!t]
    \centering
    \subfloat[]{\safeincludegraphics{0.48\textwidth}{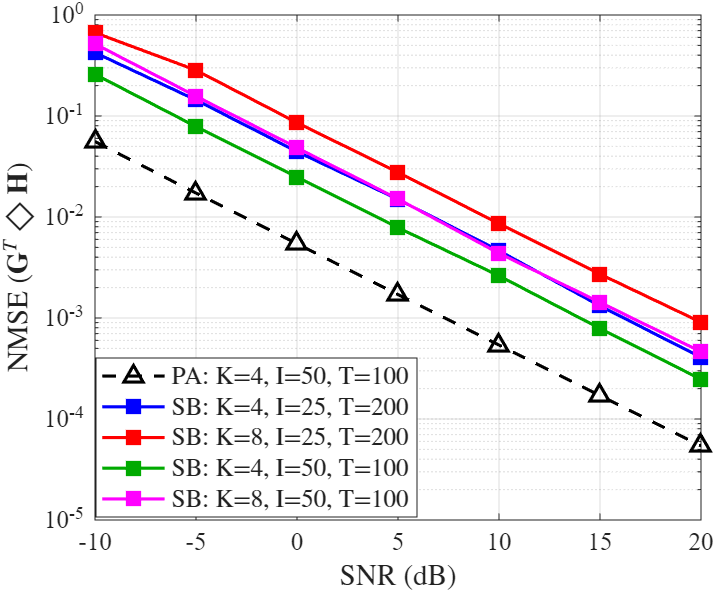}\label{fig:nmse_p1_ik}}
    \subfloat[]{\safeincludegraphics{0.48\textwidth}{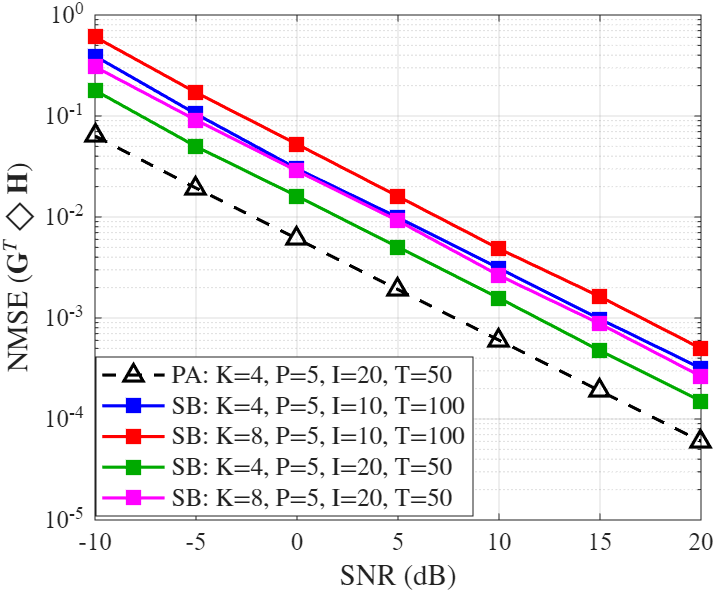}\label{fig:nmse_p2_ik}}
    \caption{NMSE sensitivity to user load and spatial-temporal resource allocation, including the pilot-assisted reference: (a) Protocol 2 for different $(K,P,T)$ settings; (b) Protocol 1 for different $(K,I,T)$ settings with $P=5$.}
    \label{fig:nmse_ik_results}
\end{figure*}

The NMSE performance of the proposed semi-blind channel estimation is further evaluated in Fig. \ref{fig:nmse_ik_results}. To assess the impact of spatial diversity and temporal resource allocation, the simulation parameters are fixed at $M=8$ active RF chains, $N=10$ total ports, and $N_r=16$ RIS elements. We analyze four different configurations for Protocol 1 (Fig. \ref{fig:nmse_ik_results}(b)) and Protocol 2 (Fig. \ref{fig:nmse_ik_results}(a)). A fundamental constraint is that the total symbol overhead is kept constant across all configurations, which is defined by the product $I \times P \times T$ for Protocol 1 and $P \times T$ for Protocol 2. As expected, increasing the number of users $K$ from 4 to 8 inherently degrades the NMSE performance in both protocols. This degradation occurs because the iterative alternating estimation procedure of the proposed PF and NPF receivers is burdened with a proportionally larger number of unknown coefficients to be jointly estimated within both the cascaded channel matrices and the symbol matrices. Furthermore, at lower SNR regimes, this expanded parameter space makes the estimation more susceptible to the noise floor.

A key observation can be extracted from a resource reallocation perspective. The results demonstrate that increasing the number of spatial diversity blocks controlled by $I$ in Protocol 1 and $P$ in Protocol 2 is more effective than simply extending the symbol period $T$. Specifically, the configurations characterized by higher spatial snapshots (e.g., the green curves with $I=20$ or $P=50$) outperform the scenarios that rely on longer temporal lengths (the blue curves with $I=10$ or $P=25$). This behavior can be explained by the fact that a higher number of spatial blocks implies more frequent variations of the antenna selection patterns ($\mathbf{S}$) and the RIS phase-shifting matrices ($\mathbf{\Phi}$). This dynamic variation enriches the spatial degrees of freedom, providing a more robust decomposition. Most notably, a trade-off is observed regarding the system scalability. The performance penalty induced by doubling the number of users ($K=8$) can be fully compensated by doubling the spatial diversity of the system. This is evident as the heavily loaded configurations with high spatial diversity (magenta curves) achieve an estimation accuracy nearly identical to, or better than, the lightly loaded configurations with low spatial diversity (blue curves). Finally, across all scenarios, the proposed semi-blind approach asymptotically approaches the Pilot-Assisted (PA) baseline (black dashed lines) at high SNR, confirming that it successfully estimates the channel components while avoiding predefined training sequences.

\begin{figure}[!t]
    \centering
    \safeincludegraphics{0.48\textwidth}{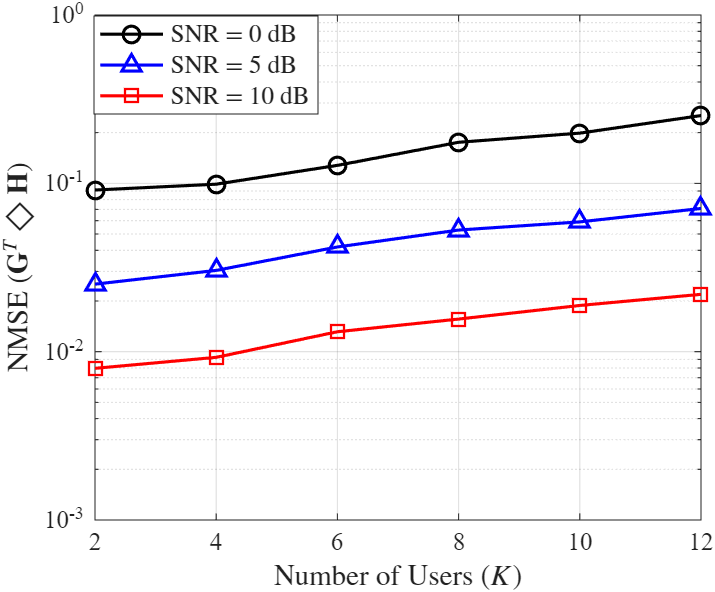}
    \caption{Scalability of the proposed semi-blind receiver with respect to user load for SNR values of 5, 10, and 15 dB.}
    \label{fig:nmse_vs_k}
\end{figure}

The scalability of the proposed tensor-based semi-blind estimation is further evaluated in Fig. \ref{fig:nmse_vs_k}, which depicts the NMSE performance as a function of the number of users $K$ for different SNR levels ($5$ dB, $10$ dB, and $15$ dB). For this experiment, we assume the configuration of Protocol 2. As expected, the results indicate that the estimation error increases monotonically with $K$ across all SNR regimes. This performance degradation is inherently linked to the semi-blind nature of the receiver. As the number of active users grows, the dimensionality of the system expands, leading to a proportionally larger number of unknown coefficients that must be jointly estimated within both the individual channel matrix $\mathbf{G}$ and the symbol matrix $\mathbf{X}$. Consequently, the overall estimation burden and the multi-user interference significantly increase. Furthermore, it can be observed that the impact of the noise floor becomes more pronounced at lower SNR values (e.g., the black curve at $5$ dB), where the proposed semi-blind receiver has fewer degrees of freedom to resolve the channel and symbol estimates. Nevertheless, despite the increased interference and the broader parameter space, the proposed iterative algorithm maintains a consistent performance trend. This behavior demonstrates that the tensor-based system scales effectively with the user load, guaranteeing reliable channel acquisition.

\begin{figure}[!t]
    \centering
    \safeincludegraphics{0.48\textwidth}{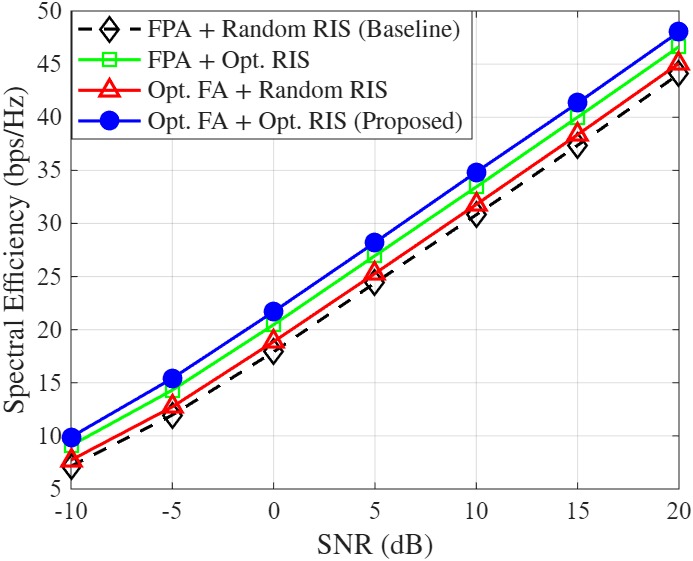}
    \caption{SE versus SNR for different RIS/FA optimization strategies in the proposed RIS-aided fluid-antenna system.}
    \label{fig:se_jfig6}
\end{figure}

The impact of RIS and FA on the spectral efficiency (SE) is investigated in Fig. \ref{fig:se_jfig6}. We consider four distinct scenarios to evaluate the SE gains provided by the proposed joint design. The baseline scenario (black dashed line) consists of Fixed Position Antennas (FPA) combined with a random RIS phase-shifting profile. As an intermediate step of improvement, we analyze the cases where either only the RIS phase-shifting values are optimized (green curve) or only the FA selection is optimized (red curve). Finally, the proposed joint optimization, which combines both FA selection and RIS phase-shifting, is depicted by the blue curve. As expected, the simultaneous optimization of the RIS and the antenna selection patterns yields the highest SE gains across the entire SNR range, confirming the synergistic advantages of integrating both technologies.

\begin{figure}[!t]
    \centering
    \safeincludegraphics{0.48\textwidth}{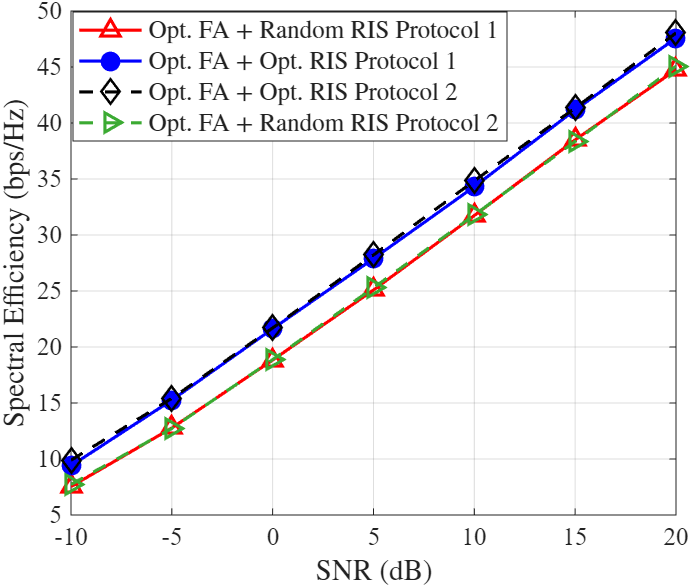}
    \caption{SE comparison between Protocols 1 and 2 under FPA + Opt. RIS and Opt. FA + Opt. RIS configurations.}
    \label{fig:se_fig7}
\end{figure}

It is worth noting the strong coupling between these two spatial domains: the optimal antenna selection matrix strictly depends on the effective channel shaped by the RIS, while the optimal RIS profile relies on the spatial channel resulting from the chosen active antennas \cite{Gil_JTSP,Asim_TCOMM_2025}. To address this non-convex joint design, we employ a robust hybrid optimization strategy. The antenna port selection is carried out via an exhaustive search over all possible discrete combinations. For each candidate FA configuration, the optimal RIS phase-shifting values are obtained through a higher-order singular value decomposition (HOSVD)-based procedure (detailed in \cite{Sokal_CAMSAP_2023}). By constructing a localized channel tensor for each candidate combination and computing its HOSVD, we extract the dominant singular vectors that represent the optimal transmission directions. This projection inherently provides the best RIS phase alignments, offering a highly effective, albeit mathematically sub-optimal, solution to the joint maximization problem. The SE comparison between Protocol 1 and Protocol 2 is subsequently presented in Fig. \ref{fig:se_fig7}. In this case, both protocols are evaluated under optimized RIS configurations, considering the cases with fixed-position antennas (FPA + Opt. RIS) and with joint fluid-antenna optimization (Opt. FA + Opt. RIS). The results show that FA optimization improves the SE for both protocols, while Protocol 1 preserves a slight performance advantage over Protocol 2 across the SNR range.

%\FloatBarrier

\section{Conclusions}
\label{sec:conclusion}

In this paper, we proposed the first joint semi-blind estimation framework for an RIS-assisted multiuser uplink communication system equipped with fluid antennas (FAs) at the base station. By leveraging the inherent multi-way structure of the received signals, we formulated two distinct signaling protocols modeled by \ac{PF} and \ac{NPF} decompositions. The proposed iterative TALS algorithms enabled the joint recovery of the individual channel matrices and data symbols with minimal pilot overhead. Our results revealed that the \ac{TTS} approach of Protocol 1, associated with the PF receiver, offers superior channel estimation accuracy and symbol recovery by more effectively exploiting the joint diversity induced by RIS phase reconfiguration and FA port selection, rapidly converging toward the perfect CSI bounds. Conversely, Protocol 2, operating under an \ac{STS} structure and associated with the NPF receiver, proved to be a highly flexible and lower-complexity alternative, particularly advantageous in scenarios with stringent temporal overhead constraints or tighter implementation requirements.
%  Extensive simulation results demonstrated that the integration of RIS and FAs yields a significant synergistic gain in spectral efficiency, substantially outperforming conventional fixed-position antenna systems. 
 Furthermore, comparisons against static RIS baselines underscored the critical role of RIS in ensuring unique joint channel and symbol estimates. Our proposed semi-blind receivers have been shown to scale effectively and accommodate an increasing number of users by strategically leveraging the spatial degrees of freedom offered by fluid antennas and RIS simultaneously. Future research directions may include the extension of this tensor framework to wideband OFDM systems and the development of low-complexity tracking algorithms for high-mobility scenarios.

\bibliographystyle{IEEEtran}
\bibliography{IEEEexample}
% \appendix
% Insert informations about the apendix here

\end{document}